\documentclass[10pt,conference]{IEEEtran}
\IEEEoverridecommandlockouts

\usepackage[
  top=0.73in,
  bottom=1.0in,
  left=0.625in,
  right=0.625in
]{geometry}

\usepackage{balance}
\usepackage{cmap}
\usepackage[utf8]{inputenc}
\usepackage[T1]{fontenc}
\usepackage{graphicx}

\usepackage{multirow, makecell}

\usepackage{amsmath}
\allowdisplaybreaks 
\usepackage{amssymb}
\usepackage{amstext}
\usepackage{amsfonts}
\usepackage{amsthm}
\usepackage{mathtools}

\usepackage{bm}
\usepackage{braket}
\usepackage{physics}
\usepackage{enumitem}

\usepackage[caption=false]{subfig}
\usepackage{overpic}

\usepackage{hyperref}
\hypersetup{colorlinks,linkcolor=blue,anchorcolor=blue,citecolor=blue,urlcolor=blue}

\usepackage{dcolumn}
\usepackage{bm}
\usepackage{array}
\usepackage{xcolor}
\usepackage{float}
\usepackage{etaremune}
\usepackage{enumitem}
\usepackage{tablefootnote}
\usepackage{url}
\usepackage{tikz}
\usepackage{qcircuit}
\usepackage{subfig}

\usepackage{subfiles}
\usepackage{etoolbox}

\theoremstyle{plain}
\newtheorem{thm}{Theorem}

\theoremstyle{definition}

\theoremstyle{remark}

\makeatother

\usepackage{orcidlink}

\begin{document}

\title{Robust Quantum Algorithmic Binary Decision-Making on Gaussian Signals
}

\author{\IEEEauthorblockN{ Aishwarya Majumdar
\orcidlink{0009-0008-2800-0455}
}
\IEEEauthorblockA{\textit{Department of Electrical and Computer Engineering, } \\
\textit{North Carolina State University, Raleigh, NC 27606, USA}\\
amajumd4@ncsu.edu}

\and

\IEEEauthorblockN{Yuan Liu
\orcidlink{0000-0003-1468-942X}
}
\IEEEauthorblockA{\textit{Department of Electrical and Computer Engineering,}
\\
\textit{Department of Computer Science,}
\\
\textit{Department of Physics and Astronomy,}
\\
\textit{North Carolina State University, Raleigh, NC 27606, USA}
\\
q\_yuanliu@ncsu.edu}
}

\maketitle

\begin{abstract}
A relevant signal in the quantum domain may manifest as a displacement or a squeezing operator in the bosonic phase space. For a real parameter $\beta$ embedded in such a Gaussian operator, the task of determining if $\beta \in [\beta_{-th}, \beta_{+th}]$ for real asymmetric thresholds $(\beta_{-th} \ne -\beta_{+th})$ is a binary decision problem. We propose a framework, the \emph{generalized quantum signal processing interferometry} (GQSPI), to solve this parameter detection problem by recasting the practical task of active binary hypothesis testing on quantum systems to a polynomial approximation problem. We achieve a small decision error probability $p_{\text{err}}$ on the order of $\mathcal{O}(\frac{1}{d}\log{(d)})$, with $d$ as the circuit depth. We analyze the protocol when (i) $\beta$ is a deterministic parameter, and (ii) when $\beta$ is drawn randomly from a known prior distribution. The GQSPI protocol is also shown to be robust under oscillator dephasing noise. We further extend our protocol from two thresholds to more general multi-threshold cases.
Overall, the proposed framework enables decision-making over arbitrary thresholds for any general Gaussian signal in a single or a few shots.
\end{abstract}

\begin{IEEEkeywords}
Quantum Detection, Binary Decision-Making, Qubit-Oscillator Hybrid Systems, Displacement, Squeezed, Generalized Quantum Signal Processing, Interferometry, Quantum Sensing
\end{IEEEkeywords}

\section{Introduction} 
\label{sec:intro}

Binary decision-making lies at the heart of several fundamental applications of information science, including sensing, communication, and computing. 
Classical detection theory \cite{wald1950statistical, van1971detection, 1988svPoor} offers a formal approach to decision problems via hypothesis testing techniques. 
This includes optimal detector designs such as the Neyman-Pearson \cite{10.1098/rsta.1933.0009, 9500a98b-1cec-3671-80dd-bc45a0de49f0} detectors that take into account cost functions and receiver operating characteristics, and the Bayesian detectors \cite{1057460, doi:10.1137/0103017, doi:10.1137/0104006} that minimize the probability of a decision error based on known information. Similarly, Helstrom and Holevo developed the early theory for quantum detection and hypothesis testing \cite{1969JSP1231H, 1968IJTP137H, HOLEVO1973337, doi:10.1137/1123048}, providing the foundations for designing optimal measurement operators and the minimum error bounds involved in discriminating quantum states, given prior probabilities. 

Practical applications of quantum detection theory include quantum sensing methods \cite{RevModPhys.89.035002, PhysRevLett.96.010401} that often deal with detection of rare astrophysical events such as dark matter interactions \cite{Bass2024,PhysRevLett.133.021801, ma2025qubitsmultilevelquantumsensing, cwx5-2n1y,
rv43-54zq} or gravitational waves \cite{ligo2015squeezing, bothwell2022resolving, ligo2016observation}, where the presence of minute signals must be distinguished from noise. For quantum communication and computing \cite{von1955mathematical, 10.5555/3240076, wilde2013quantum, nielsen2000quantum}, the task of determining an encoded message transmitted in presence of noise is central to the design of optimal decoders and to achieving the channel capacity \cite{PhysRevLett.66.1119, PhysRevA.54.1869, PhysRevA.51.2738}. 

Properties unique to quantum systems such as entanglement and superposition have enabled the performance of parameter estimation techniques to go beyond the classical limits, also known as the \emph{standard quantum limit} (SQL), and achieve Heisenberg-limited (HL) scaling. Likewise, for binary detection problems, quantum protocols, such as quantum illumination, \cite{doi:10.1126/science.1160627,PhysRevLett.101.253601} have shown improvements over classical methods. In many practical scenarios, the goal is to determine whether a parameter lies within one or more specified regions of the parameter space. However, such decision problems in the quantum domain are far more delicate because quantum signals are often encoded in an unknown quantum state or process, buried under classical environmental noise, where the quantum states can undergo rapid decoherence and decay during the detection process. Although several techniques have been proposed to overcome these hurdles \cite{PhysRevLett.130.070802, Kuniyil_2022, Myatt2000, PhysRevB.77.174509}, a systematic approach offering provable improvements in decision accuracy for active detection of quantum Gaussian signals is largely missing. 

Important classes of quantum detection problems include asymmetric and multi-thresholding problems in systems with asymmetric or nonlinear responses, where the physical transduction of a gaussian signal induces an asymmetric mapping in phase space, leading to decision boundaries that cannot be efficiently captured by symmetric thresholding methods. Having such flexible detection capabilities enable robust general decision-making in the presence of noise and uncertainty, motivating the need for detection frameworks that go beyond simple binary symmetric thresholds.

In this work, we establish a framework for solving quantum detection problems in which an underlying Gaussian signal parameter may be deterministic or a random variable with a known prior distribution, with the goal of determining whether the parameter lies between two given real thresholds (say, $\beta_{-th}$, and $\beta_{+th}$). The framework incorporates non-Gaussian resources by applying \emph{generalized quantum signal processing} (GQSP) algorithm \cite{PRXQuantum.5.020368} on hybrid qubit-bosonic oscillator systems, utilizing the infinite-dimensional Hilbert space of the oscillator to sense continuous-variable Gaussian signal parameter and entangling it with the discrete qubit measurement outcome as the binary decision result. Our approach, \emph{generalized quantum signal processing interferometry} (GQSPI) can achieve asymmetric thresholding detection and more general multi-thresholding cases,
improving upon prior art using \emph{quantum signal processing interferometry} (QSPI) \cite{SinananSingh2024singleshotquantum} which can only handle symmetric thresholding problems ($\beta_{-th} = -\beta_{+th}$) due to the limited expressivity of the \emph{quantum signal processing} (QSP) algorithm \cite{low2016methodology, low2019hamiltonian, PhysRevLett.118.010501}.

Additionally, we show that the GQSPI protocol is essentially robust against oscillator dephasing noise for displacement signal detection. Further extending the work to squeezing signals, we discuss how the protocol offers the possibility of evaluating the decision problem with the multi-resolution capabilities for detecting squeezed signals across asymmetric thresholds.
Such a scenario can be useful in detecting whether the squeezing strength lies within an optimal interval, which is critical in protocols where both insufficient and excessive squeezing can degrade performance \cite{9542141}. 


The remainder of the paper is organized as follows. Section~\ref{sec:bg} provides the necessary background on the notations used throughout the paper, followed by a brief introduction to the \emph{quantum signal processing} algorithm.
The main problem formulation and proposed technical approach is discussed in Sec.~\ref{sec:mp_ta}. Extensions of the GQSPI framework to multi-threshold detection, stochastic $\beta$, oscillator dephasing noise, and quadratic Gaussian signal detection are analyzed in Sec.~\ref{sec:extensions-gqspi}. Finally, numerical simulation results for both single-band and multi-band threshold detection are presented in Sec.~\ref{sec:results}. We conclude the work with a brief discussion on the future directions in Sec.~\ref{sec:disc}.


\section{Background} \label{sec:bg}
\subsection{Notations} \label{sec:bg-notations}

Throughout the paper, we refer to the canonical position and momentum operators of harmonic oscillators as $\hat{x}$ and $\hat{p}$, satisfying $[\hat{x}, \hat{p}] = i$, assuming $\hbar = 1$, mass $m=1$, and frequency $\omega=1$. $a$ and $a^{\dagger}$ are the bosonic annihilation and creation operators satisfying $[a, a^{\dagger}] = 1$ with $\hat{n} = a^{\dagger}a$ being the photon number operator. Using `Standard units' \cite{4rf7-9tfx}, these operators are related as $a = \frac{1}{\sqrt{2}}(\hat{x} + i\hat{p}), a^{\dagger} = \frac{1}{\sqrt{2}}(\hat{x} - i\hat{p})$ and $\hat{x} = \frac{1}{\sqrt{2}}(a + a^{\dagger}), \hat{p} =-i\frac{1}{\sqrt{2}}( a - a^{\dagger})$. 
The vacuum or ground state of a harmonic oscillator is denoted by $\ket{0}_{osc}$ which is a Gaussian state. 
$I_{osc}$ represents the identity operation on the oscillator. 

We denote the qubit ground state and the excited state by $\ket{\downarrow}$ and $\ket{\uparrow}$, respectively. $\sigma_x, \sigma_y, \sigma_z$ are the qubit Pauli operators.

\subsection{Tools} \label{sec:bg-tools}
Of the several advances in quantum algorithms in the past decade, Quantum Signal Processing (QSP) \cite{low2016methodology, low2019hamiltonian, PhysRevLett.118.010501} and its generalized form, GQSP \cite{PRXQuantum.5.020368} have proven to be exceptionally versatile algorithms in terms of their applicability to solutions for a wide variety of quantum problems, including hamiltonian simulation, quantum search, and factoring \cite{MRTC21}. The standard form of QSP involves interleaving two types of qubit rotation operators repeatedly, namely a signal operator encoding a parameter of interest, and a signal processing operator. The resultant unitary at the end of this protocol realizes a polynomial transformation on the parameter. This process of interleaving two different families of quantum operators forms the backbone of the binary decision approach presented in the following section.


\section{Main Problem and Technical Approach}\label{sec:mp_ta}

We outline the main problem in Sec.~\ref{sec:mp_ta-problem}, followed by the technical approach, and key results of this work in the subsequent subsections. We introduce GQSP for hybrid qubit–oscillator systems (Theorem~\ref{thm-1}) in Sec.~\ref{sec:mp_ta-hybrid-gqsp}, which forms the foundation for the GQSPI protocol presented in Sec.~\ref{sec:mp_ta-gqspi-thm} (Theorem~\ref{thm-2}). Error scaling of the GQSPI framework is analyzed in Sec.~\ref{sec:results-error}.

\subsection{Main Problem} \label{sec:mp_ta-problem}

The Gaussian signal $S_\beta$, defined as follows 
\begin{align}
    S_\beta = I \otimes e^{i\beta \hat{p}^{k}} =
    \begin{bmatrix}
        e^{i \beta \hat{p}^{k}}  & 0 \\
        0 & e^{i \beta \hat{p}^{k}}
    \end{bmatrix},~ \beta \in \mathbb{R},~ k=\{1,2\} \label{s-beta}.
\end{align}
is assumed to interact with the qubit-bosonic oscillator system. For $k=1$, $S_\beta$ is a displacement signal which applies a position change of $\beta$ to the oscillator state. When $k=2$, $S_\beta$ is a quadratic Gaussian signal similar to squeezing signals. 

With the goal to determine whether the value of $\beta$ lies within the range $[\beta_{-th}, \beta_{+th}]$,
the GQSPI protocol prepares the qubit–oscillator system so as to minimize the overall binary decision error. The main problem we solve in this work is:

\textbf{Main Problem:} \emph{Given thresholds $\beta_{-th}$,  $\beta_{+th} \in \mathbb{R}$, determine whether the parameter $\beta \in \mathbb{R}$, embedded in a Gaussian operator $S_{\beta}$ as shown in Eq.~\eqref{s-beta}
in the qubit-oscillator space, 
lies within the range $[\beta_{-th}, \beta_{+th}]$ or outside of it, with a small decision error probability $p_{\text{err}}$, using a quantum circuit constructed using qubit rotation gates and qubit-oscillator entangling gates.}

We assume that the qubit is in ground state $\ket{\downarrow}$, and the oscillator is in the vacuum state $\ket{0}_{osc}$ at the beginning of the protocol. Following the GQSPI protocol presented in Theorem~\ref{thm-2}, the qubit is measured along the Pauli-Z basis, where the measurement outcome $\downarrow$ signifies that the parameter $\beta$ is in the range $[\beta_{-th}, \beta_{+th}]$, while measuring $\uparrow$ implies that $\beta$ does not belong to this range. We denote $P(M = \downarrow| \beta)$ as the probability of measuring $\downarrow$ on the qubit for a given value of $\beta$.

\subsection{ GQSP Theorem for Qubit-Oscillator Systems} \label{sec:mp_ta-hybrid-gqsp}

Consider the following single-qubit rotation parametrized by angles $\{\theta, \phi, \lambda\}$
\begin{align}
    R(\theta, \phi, \lambda) 
    &= \label{eq:r}
    \begin{bmatrix} 
        e^{i(\lambda+ \phi)}\cos{\theta} & e^{i\phi}\sin{\theta} \\
        e^{i\lambda}\sin{\theta} & 
        -\cos{\theta}
    \end{bmatrix} 
\end{align}
and an entangling operator between the qubit and the oscillator
\begin{align*}
    W_z(\hat{\omega}) = \begin{bmatrix}
        \hat{\omega}(\hat{x}, \hat{p}) & 0 \\
        0 & \hat{\omega}^{-1}(\hat{x}, \hat{p})
    \end{bmatrix},
\end{align*}
which applies operator $\hat{\omega}(\hat{x}, \hat{p})$ to the oscillator state if the qubit state is $\ket{\downarrow}$ and $\hat{\omega}^{-1}(\hat{x}, \hat{p})$ if the qubit state is $\ket{\uparrow}$. As a building block, we state the \emph{generalized quantum signal processing} (GQSP) for hybrid qubit-oscillator systems as follows:
\begin{thm}\label{thm-1}
    A quantum circuit on hybrid qubit-oscillator system parameterized by the set of angles $\vec{\theta} = \{\theta_0, \theta_1 \cdots \theta_d\}, ~\vec{\phi} = \{\phi_0, \phi_1 \cdots \phi_d\}$, and $\lambda_0$ realizes a block-encoding of a degree-$d$ complex Laurent polynomial transformation $P(\hat{\omega})$ on an oscillator operator $\hat{\omega}(\hat{x}, \hat{p})$ using the following sequence given by $G_{\vec{\theta}, \vec{\phi}, \lambda_0}(\hat{\omega})$:
\begin{align}
    &\nonumber
    G_{\vec{\theta}, \vec{\phi}, \lambda_0}(\hat{\omega}) 
    \\ 
    \nonumber
    &= \left(\prod_{i=1}^{d} (R(\theta_i, \phi_i, 0)\otimes I_{osc})W_z(\hat{\omega})\right)(R(\theta_0, \phi_0, \lambda_0)\otimes I_{osc}) ,
    \\ 
    &= \begin{bmatrix}
        P_d(\hat{\omega}) &  -Q^{\dagger}_d(\hat{\omega})
        \\
        Q_d(\hat{\omega}) 
        & P^{\dagger}_d(\hat{\omega})
    \end{bmatrix},
\end{align}
such that,
\begin{itemize}
    \item $P_d(\hat{\omega})$ and $Q_d(\hat{\omega})$ are both degree-$d$ polynomials in $\hat{\omega}$, 
    \item $|P_d(\hat{\omega})|^2 + |Q_d(\hat{\omega})|^2 = 1 ~\forall~\hat{\omega}$, 
    \item $P_d(\hat{\omega})$ and $Q_d(\hat{\omega})$ are of the form,
    \\
    $P_d(\hat{\omega}) = \sum_{n=-d}^{d} p_n \hat{\omega}^n,
    \quad
    Q_d(\hat{\omega}) = \sum_{n=-d}^{d} q_n \hat{\omega}^n$ where
    $n = \{-d, -d+2, \cdots, d-2, d\}$, and $p_n, q_n \in \mathbb{C}$.
\end{itemize}
\end{thm}
Similar form of GQSP for hybrid qubit-bosonic oscillator systems was also independently obtained by \cite{hong2025oscillatorqubitgeneralizedquantumsignal}.
The nature of polynomials and relationships between polynomial coefficients can be found Appendix \ref{sec:appndx-GQSP-Poly-Forms}.

\subsection{ Generalized Quantum Signal Processing Interferometry} \label{sec:mp_ta-gqspi-thm}

\begin{figure}[ht]
    \centering
    \includegraphics[width=1\linewidth]{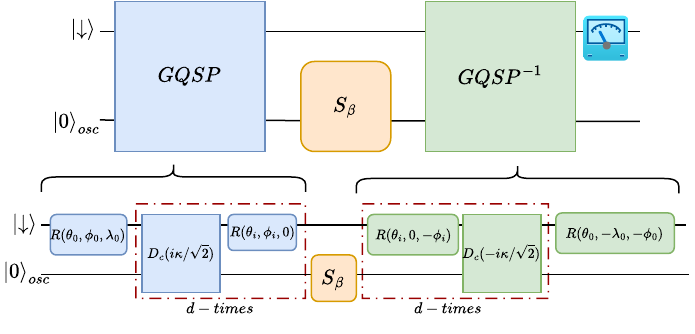}
    \vspace*{-1\baselineskip}
    \caption{Quantum circuit for Generalized Quantum Signal Processing Interferometry (GQSPI). Note that for the qubit rotation gate in Eq. \eqref{eq:r}, $R^{\dagger}(\theta, \phi, \lambda) = R(\theta, -\lambda,  -\phi)$.}
    \label{fig:gqspi-setup}
\end{figure}

Let the qubit-oscillator entangling operator $W_z(\hat{\omega})$ be a conditional displacement gate on the oscillator, conditioned on the qubit state, given by, 
\begin{align}
    \mathcal{D}_c(\alpha) = e^{(\alpha a^{\dagger} - \alpha^* a)\sigma_z} 
\end{align}
where $a$ and $a^{\dagger}$ are the bosonic annihilation and creation operators on the oscillator as noted in Sec. \ref{sec:bg-notations}.
In particular, for $\alpha = i \kappa / \sqrt{2}$ where $\kappa \in \mathbb{R}$, the $\mathcal{D}_c(\alpha)$ operator imparts a momentum kick to the oscillator by $+\kappa$ if the qubit state is $\ket{\downarrow}$ and $-\kappa$ if the qubit state is $\ket{\uparrow}$,
\begin{align}
    \mathcal{D}_c(i \kappa / \sqrt{2}) = e^{i \kappa \hat{x} \hat{\sigma}_z} = \begin{bmatrix}
        e^{i \kappa \hat{x}} & 0 
        \\
        0 & e^{-i \kappa \hat{x}}
    \end{bmatrix}
    \label{Dc-def}.
\end{align}
It can be observed that the $\mathcal{D}_c(i \kappa / \sqrt{2})$ gate is periodic in the position eigen state $\ket{x}_{osc}$ with periodicity $T_{\mathcal{D}} = \frac{2\pi}{\kappa}$ in the bosonic phase-space.

For the interferometry protocol, we sandwich the displacement signal containing operator $S_{\beta}$ ($k=1$) shown in Eq.~\eqref{s-beta}, between a degree-$d$ GQSP sequence and its inverse sequence, 
\begin{align}
     U(\beta, \kappa)
     &= \label{eq:U-def}
     G^{-1}_{d,\vec{\theta}, \vec{\phi}, \lambda_0}(\hat{x}) ~S_{\beta}~ G_{d,\vec{\theta}, \vec{\phi}, \lambda_0}(\hat{x})
\end{align}
where the GQSP sequence consists of qubit rotation gates $R(\theta, \phi, \lambda)$ and qubit-oscillator entangling gate $\mathcal{D}_c(i \kappa / \sqrt{2})$ as follows:
\begin{align}
     G_{d,\vec{\theta}, \vec{\phi}, \lambda_0}(\hat{x}) 
     &= \nonumber
     \left[(R(\theta_d, \phi_d, 0)\otimes I_{osc})D_c(i\kappa/\sqrt{2})\right] 
     \\
     & \nonumber
     \cdots \times\left[(R(\theta_1, \phi_1, 0)\otimes I_{osc})D_c(i\kappa/\sqrt{2})\right] 
     \\
     & \label{eq:gsqp-expanded}
     \times
     (R(\theta_0, \phi_0, \lambda_0)\otimes I_{osc}).
\end{align}
From the hybrid GQSP Theorem \ref{thm-1}, $G_{\vec{\theta}, \vec{\phi}, \lambda_0}(\hat{x})$ can be expressed as, 
\begin{align*}
    G_{\vec{\theta}, \vec{\phi}, \lambda_0}(\hat{x}) 
    &=
    \begin{bmatrix}
         P_{d}(\hat{x}) &  -Q^{\dagger}_{d}(\hat{x})
         \\
         Q_{d}(\hat{x}) &  P^{\dagger}_{d}(\hat{x})
    \end{bmatrix}
\end{align*}
where, $P_{d}(\hat{x}) = \sum_{n=-d}^{d} p_n e^{i n\kappa \hat{x}} $ and $Q_{d}(\hat{x}) = \sum_{n=-d}^{d} q_n e^{i n\kappa \hat{x}}$ with $p_n, q_n \in \mathbb{C}$.

By choosing appropriate phase angles $\vec{\theta} = \{\theta_0, \theta_1 \cdots \theta_d\}, ~\vec{\phi} = \{\phi_0, \phi_1 \cdots \phi_d\}$, and $\lambda_0$, the probability of measuring the qubit in state $\ket{\downarrow}$ for given $\beta$, denoted by $P(M = \downarrow| \beta)$ can be made close to 1 for $\beta \in [\beta_{-th}, \beta_{+th}]$ and close to 0 when $\beta$ is outside this range. The probability of measuring $\ket{\downarrow}$ for a given $\beta$ is obtained to be,
\begin{align}
    P(M = \downarrow| \beta) 
    &= \nonumber
   \langle 0|_{osc} U^{\dagger}(\beta, \kappa)_{00}U(\beta, \kappa)_{00}|0 \rangle _{osc}
   \\
   &
   = \label{eq:csbeta}
   \sum_{s = -d}^{d} c_s e^{i(2\kappa) \beta s},
   \end{align} 
   with,
   \begin{align}
   &c_s 
    = \nonumber
      \sum_{n, m, r = -d}^{d}
    \left( p_n p^{*}_{m} +  q_n q^{*}_{m} \right)
    \\
    & 
    \label{eq:qspi_eq_20}
    \times
    \left( p^{*}_{n + 2s} p_{m + 2r} +  q^{*}_{n + 2s} q_{m + 2r} \right) 
     e^{- \kappa^2 (r - s)^2}
\end{align}
where $s,r,m,n = \{-d, -d+2, \cdots, d-2, d\}$.
$U(\beta, \kappa)_{00}$ in Eq. \eqref{eq:csbeta} denotes the top-left element of the operator $U(\beta, \kappa)$ which can be expressed as,
\begin{align}
    U(\beta, \kappa)_{00} 
    &= \label{eq:U00}
    \sum_{n,m=-d}^{d} \left( p^{*}_n p_m +  q^{*}_n q_m \right)e^{-i n\kappa \hat{x}} e^{i\beta \hat{p}} e^{i m \kappa \hat{x}}.
\end{align}
It is interesting to note that $P(M = \downarrow| \beta)$ is a polynomial in $e^{i(2\kappa) \beta}$, and has a periodicity of $T_P = \frac{\pi}{\kappa}$. Therefore, it is necessary to choose $\kappa$ appropriately such that $\beta \in \left(-\frac{\pi}{2\kappa}, \frac{\pi}{2\kappa}\right)$ so as to cover the entire range of $\beta$.
$P(M = \downarrow | \beta)$ can be an asymmetric ($c_s \ne c_{-s}$) or symmetric ($c_s = c_{-s}$) function of $\beta$ depending on the choice of phase angles, unlike in QSPI \cite{SinananSingh2024singleshotquantum} where $P(M = \downarrow | \beta)$ is always a symmetric function, thus limiting its usefulness for asymmetrically distributed parameter detection.

In the absence of any displacement signal, $\beta = 0$, thus no displacement is applied to the oscillator, and the GQSP and its inverse sequence cancel out, implying no change in the initial qubit state. This implies that $\sum_{s = -d}^{d} c_s = 1$ for $\beta = 0$. For cases where $0\notin [\beta_{-th}, \beta_{+th}]$, $P(M = \downarrow| \beta) = 1$ at $\beta=0$ can be treated as a point of exception.

\subsection{Error Scaling and Circuit Depth}\label{sec:results-error}
\begin{figure}[ht]
    \centering
    \includegraphics[width=1\linewidth]{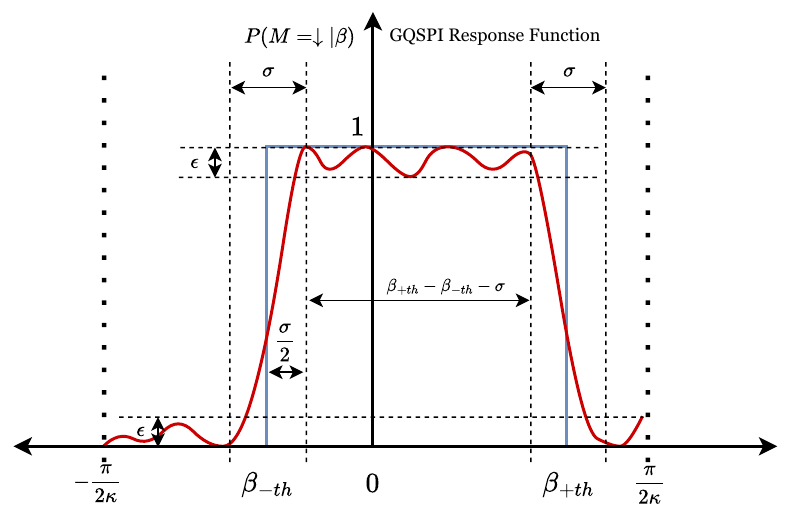}
     \vspace*{-1.5\baselineskip}
    \caption{Error analysis of the GQSPI protocol.}
    \label{fig:err-analysis}
\end{figure}
Now that the qubit measurement probability response function has been expressed as polynomial in Eq. \eqref{eq:csbeta}, the error density of the protocol can be viewed as the error in approximating the ideal function,
\begin{align}
    P_{\text{ideal}} = \label{eq:p_ideal}
    \begin{cases}
    1 , \quad \beta_{-th} \leq \beta \leq \beta_{+th}
    \\
    0, \quad \text{otherwise}
\end{cases}.
\end{align}
The probability of decision error density $p_{\text{err}}(\beta_{-th}, \beta_{+th}, \kappa)$ of the protocol can be expressed as,
\begin{align}
    &p_{\text{err}}(\beta_{-th}, \beta_{+th}, \kappa) \nonumber
    \\
    &
    = \nonumber
    \frac{\kappa}{\pi} 
    \int_{-\frac{\pi}{2\kappa}}^{\frac{\pi}{2\kappa}} 
    |P_{\text{ideal}} - P(M = \downarrow | \beta)|d\beta
    \\
     &= \label{eq:P-err-form}
      \frac{ (\beta_{+th} - \beta_{-th})\kappa}{\pi}
    +
    \sum_{s = -d}^{d} c_s H_s, 
    \\
    &
    \text{where,} \nonumber
    \\
    &H_s = \text{sinc}(\pi s) + i \frac{e^{i( 2\kappa \beta_{+th}) s} - e^{i( 2\kappa \beta_{-th}) s}}{\pi s}.
\end{align}
As shown in Fig. \ref{fig:err-analysis}, assume that the error tolerance of the protocol is $\epsilon$ everywhere expect for $\beta \in [\beta_{-th}-\frac{\sigma}{2}, \beta_{-th}+\frac{\sigma}{2}] \cup [\beta_{+th}-\frac{\sigma}{2}, \beta_{+th}+\frac{\sigma}{2}]$ which are considered as the transition regions of width $\sigma$ around each of the thresholds.
Then the total probability of decision error denoted by $p_{\text{err}}$ can be expressed as,
\begin{align}
    p_{\text{err}} 
    =
    \epsilon \left( \frac{\pi}{\kappa} - 2\sigma \right) + \frac{\sigma}{2}
\end{align}
where we assume that $\epsilon$ and $\sigma$ are of the same order of magnitude. Using the polynomial approximation of a rectangular function result obtained in \cite{low2017hamiltoniansimulationuniformspectral, SinananSingh2024singleshotquantum}, the degree of the polynomial $d$ is related to $p_{\text{err}}$ as, 
\begin{align}
    d &\propto \frac{1}{\kappa p_{\text{err}}}\log{(\kappa p_{\text{err}})}
    \quad
    \text{or equivalently,} \label{eq:degree-perr}
    \quad
    p_{\text{err}} \propto \frac{1}{\kappa d}\log{(d)}.
\end{align}
This brings us to stating the following theorem:
\begin{thm}\label{thm-2}
    An interferometry protocol obtained by applying a degree-$d$ hybrid-GQSP before and after the displacement signal $S_\beta$ (with $k=1$) acts on an oscillator, realizes the probability of measuring the qubit in ground state $P(M = \downarrow | \beta) = \sum_{s = -d}^{d} c_s e^{i(2\kappa) \beta s}, \kappa\in\mathbb{R}$, achieving arbitrary polynomials w.r.t. $e^{i2\kappa\beta}$, thus improving over QSPI \cite{SinananSingh2024singleshotquantum} which can only realize symmetric (even) functions. Such polynomial transformation produces a desirable interference on the quantum state of the system suitable for binary decision-making with probability of decision error $p_{\text{err}}  \propto \mathcal{O}(\frac{1}{\kappa d}\log{d})$. 
\end{thm}

Numerical optimization and analytical methods established in several of the QSP angle finding techniques \cite{chao2020findinganglesquantumsignal, Haah2019product, laneve2025generalizedquantumsignalprocessing} can be used to obtain $\vec{\theta} = \{\theta_0, \theta_1 \cdots \theta_d\}, ~\vec{\phi} = \{\phi_0, \phi_1 \cdots \phi_d\}$, and $\lambda_0$ that create the optimal detection state on the qubit-oscillator system. 

The detailed treatment of the GQSPI protocol is presented in Appendix~\ref{sec:appndx-GQSPI}
with derivations for $P(M = \downarrow | \beta)$ and $p_{\text{err}}(\beta_{-th}, \beta_{+th}, \kappa)$.

\section{Extensions of GQSPI Framework} \label{sec:extensions-gqspi}
In this section, we discuss the extensions of the GQSPI framework to multi-threshold cases in Sec.~\ref{sec:mp_ta-multi-thresh}, explore the stochastic setting in Sec.~\ref{sec:mp_ta-beta-rv}, and examine the robustness of the protocol under dephasing noise in Sec.~\ref{sec:mp_ta-dephasing}. In Sec.~\ref{sec:mp_ta-beta-squeezing}, we analyze the protocol for quadratic signals.

\subsection{Displacement Signal Detection across Multiple Thresholds} \label{sec:mp_ta-multi-thresh}
We assume $\beta$ is required to be detected for $m$ number of non-overlapping bands of thresholds indexed by pairs 
\begin{align}
    \{\vec{\beta}_{-th}, \vec{\beta}_{+th}\} &= \nonumber
    \{ \{\beta_{1,-th}, \beta_{1,+th}\}, \{\beta_{2,-th}, \beta_{2,+th}\}, 
    \\
    & 
    \cdots, \{\beta_{m,-th}, \beta_{m,+th}\}\}.
\end{align}
Then, the expression for $p_{\text{err}}(\{\vec{\beta}_{-th}, \vec{\beta}_{+th}\}, \kappa)$ is obtained to be,
\begin{align}
    &p_{\text{err}}(\{\vec{\beta}_{-th}, \vec{\beta}_{+th}\}, \kappa) 
    \nonumber
    \\
    &
    = \nonumber
    \frac{\kappa}{\pi} \sum_{j=1}^{m}(\beta_{j,+th} - \beta_{j,-th})
    + \nonumber
    \sum_{s = -d}^{d} c_s 
     \text{sinc}(\pi s)
     \\
     &+ \label{eq:general-multi-thresh-perr}
      \sum_{s = -d}^{d} c_s 
     \left[
     (\frac{i}{\pi s}) \sum_{j=1}^{m} \left(
     e^{i( 2\kappa \beta_{j,+th}) s} - 
     e^{i( 2\kappa \beta_{j,-th}) s}\right)
     \right].
\end{align}
Thus, the single band of threshold case is easily generalizable to multiple bands of thresholds. Simulation results for a simple case of two bands of thresholds is shown in Fig.~\ref{fig:double-window}, and calculation details are presented in Appendix \ref{sec:multi-threshold}.

\subsection{Displacement Signal Detection for Stochastic Parameter} \label{sec:mp_ta-beta-rv}
Oftentimes the parameter of interest is a random variable. Let $\beta$ follow a probability density function $f_{\beta}(\mu_{\beta}, \sigma^2_{\beta})$ with  mean $\mu_{\beta}$, and variance $\sigma^2_{\beta}$. In this case, the decision error density transforms into an expectation value
\begin{align}
    &\nonumber
    p_{\text{err}}(\beta_{-th}, \beta_{+th}, \kappa, \sigma_{\beta}, \mu_{\beta})
     = 
     \frac{\kappa}{\pi}
     \mathbb{E} \left[
    |P_{\text{ideal}} - P(M = \downarrow | \beta) |\right ],
    \\
    &\text{where, } \nonumber
    \\
    &
    \mathbb{E}[ P(M = \downarrow | \beta) ] = \sum_{s = -d}^{d} c_s \left[\int_{-\infty}^{\infty}e^{i(2\kappa)\beta s}
      f_{\beta}(\mu_{\beta}, \sigma^2_{\beta}) d\beta \right].
\end{align}
Further, when actually using the GQSPI protocol, since the only change required is to adjust the decision error density $p_{\text{err}}(\beta_{-th}, \beta_{+th}, \kappa)$, therefore, the optimization algorithm minimizing the decision error density can accommodate the prior information of $\beta$ to realize an appropriate polynomial transformation.
The particular case of $\beta$ following the Gaussian probability density is presented in Appendix \ref{sec:appndx-gaussian-beta}.
The current treatment can also be recast for deterministic $\beta$ buried in classical random noise with a known distribution.

\subsection{GQSPI with Dephasing Noise on the Oscillator} 
\label{sec:mp_ta-dephasing}
Similar to qubit decoherence errors, the oscillator also suffers from dephasing. We assume such a dephasing error occurs during the application of each conditional displacement gate, given by the dephasing operator $R_{osc}(\gamma) = I \otimes e^{-i\gamma \hat{n}}$. The parameter $\gamma$ could be a random variable drawn from a known distribution, but is usually a fixed value in the form of a coherent error occurring during the duration of the protocol (for shorter circuits).
 This will change $\mathcal{D}_c(i \kappa / \sqrt{2})$ into a noisy version:
 \begin{align}
        \mathcal{D}_c(\alpha) &= \nonumber
        R_{osc}^\dagger(\gamma) \mathcal{D}_c(i \kappa / \sqrt{2}) R_{osc}(\gamma) 
        \\
        &=
        \begin{bmatrix}
        e^{i \kappa (\hat{x} \cos\gamma  + \hat{p} \sin\gamma) } & 0 \\
        0 & e^{-i \kappa (\hat{x} \cos\gamma  + \hat{p} \sin\gamma) }
    \end{bmatrix}.
    \end{align}
Then the GQSPI sequence with dephasing error occurring for each iteration is of the form,
\begin{align}
    G_{\vec{\theta}, \vec{\phi}, \lambda_0}(\omega) 
    &= \nonumber
    \left(\prod_{i=1}^{d} \Bigl(R(\theta_i, \phi_i, 0)\otimes I_{osc}\Bigr) \mathcal{D}_c(i \kappa / \sqrt{2}) R_{osc}(\gamma_i) \right)
    \\
    & \label{eq:dephasing-gqspi-formulation}
    \times \Bigl(R(\theta_0, \phi_0, \lambda_0)\otimes I_{osc}\Bigr)
    \\
    &= \nonumber
    R_{osc}(\Gamma_i)  \left(\prod_{i=1}^{d} (R(\theta_i, \phi_i, 0)\otimes I_{osc}) \mathcal{D}_c(\alpha'_i) \right)
    \\
    &
    \times
    \Bigl(R(\theta_0, \phi_0, \lambda_0)\otimes I_{osc} \Bigr) 
    \end{align}
where,
\begin{align}
    &\alpha'_i = \frac{i\kappa \cos\Gamma_i + \kappa \sin\Gamma_i}{\sqrt{2}} = \frac{\kappa}{\sqrt{2}}e^{i(\frac{\pi}{2} -\Gamma_i)}, 
    \\
    &\text{and}\quad 
    \Gamma_i = \sum_{j=1}^{i} \gamma_j.
\end{align}
The dephasing noise rotates each of the oscillator displacements conditioned on the qubit by a varying amount, accumulating all the random rotation angles from the previous iterations.
For such a GQSP set up, the general form of the polynomial 
$P_d(\vec{\gamma})$, where $\vec{\gamma}= \{\gamma_1,\gamma_2 \cdots, \gamma_d\}$, can be expressed as, 
\begin{align}
    P_d(\vec{\gamma}) 
    &= \nonumber
    \sum_{\textbf{s} \in \{-1, 1\}^d} p_{\textbf{s} } \mathcal{D}\left(
    \sum_{k=1}^{d}s_k\alpha'_k 
    \right)
    \\
    & \label{eq:gen-poly-dephasing}
    \times
    \exp{
    \left(
    -i\frac{\kappa^2}{2}
    \sum_{1 \leq l < k \leq d } s_k s_l
    \sin(\Gamma_k - \Gamma_l)
    \right)}, 
\end{align}
where,
\begin{align*}
    &s_k \in \{-1, 1\}~\forall~k~\in \{1,2, \cdots, d\},
    \quad
   \text{and} \quad
    p_{\textbf{s}} \in \mathbb{C}.
\end{align*}
$Q_d(\vec{\gamma})$ is also of similar form.
Thus, the polynomial $P$ can be viewed as a linear combination of rotated oscillator displacement operators with varying magnitude and direction.

The overall unitary acting on the qubit-oscillator system taking into account the dephasing noise for the entire interferometry protocol is given by,
 \begin{align}
     &U(\beta, \kappa, \vec{\gamma}) 
     = 
     \nonumber G^{-1}_{d,\vec{\theta}, \vec{\phi}, \lambda_0}(\vec{\gamma}) ~S_{\beta}~ G_{d,\vec{\theta}, \vec{\phi}, \lambda_0}(\vec{\gamma})
     \\
     &= \label{eq:u-deph}
     \begin{bmatrix}
         P^{\dagger}_d(\vec{\gamma})& Q^{\dagger}_d(\vec{\gamma})
         \\
         * & *
     \end{bmatrix}
     I \otimes \mathcal{D}\left( -\frac{\beta}{\sqrt{2}}e^{i\Gamma_d} \right)
     \begin{bmatrix}
         P_d(\vec{\gamma})& *
         \\
         Q_d(\vec{\gamma})& *
     \end{bmatrix}.
 \end{align}
The $P(M = \downarrow| \beta)$ obtained for the dephasing noise scenario is of the form,
\begin{align*}
    P(M = \downarrow| \beta) 
   &=
   \sum_{\textbf{s}, \textbf{s'}, \textbf{r}, \textbf{r'}\in \{-1, 1\}^d}  
    (p^*_{\textbf{s}}p_{\textbf{s'} } + q^*_{\textbf{s} }q_{\textbf{s'} })
    (p_{\textbf{r}}p^*_{\textbf{r'} } + q_{\textbf{r} }q^*_{\textbf{r'} })
    \\
    &
    \times
    e^{i\Theta_{\textbf{s}, \textbf{s'}, \textbf{r}, \textbf{r'}}}
    e^{-|\alpha_{\textbf{s}, \textbf{s'}, \textbf{r}, \textbf{r'}}|^2/2}
    e^{i(\kappa\beta)\Lambda_{\textbf{s}, \textbf{r}}} 
\end{align*}
where $e^{i\Theta_{\textbf{s}, \textbf{s'}, \textbf{r}, \textbf{r'}}}$ and $ e^{-|\alpha_{\textbf{s}, \textbf{s'}, \textbf{r}, \textbf{r'}}|^2/2}$ are overall phase and amplitude factors, respectively, obtained from composing the scrambled conditional displacement gates, and $e^{i\kappa\beta \Lambda_{\textbf{s}, \textbf{r}}} $ is comparable to the factor $e^{i 2\kappa \beta s}$ obtained previously in Eq.~\eqref{eq:csbeta}, and \eqref{eq:qspi_eq_20}. Detailed expressions of the parameters can be found in Appendix \ref{sec:appnx-gqspi-deph}.

It can be shown that for small dephasing rates $\gamma$ (with same $\gamma$ in every iteration), $P(M = \downarrow| \beta)$ has a rescaling factor of $(1 - \Omega_1\gamma^2 - \Omega_2\beta \gamma^3 + \mathcal{O}(\gamma^4))$ for $\Omega_1, \Omega_2\in\mathbb{R}$, which implies that $P(M = \downarrow| \beta)$ remains close to 1 up to a 2nd-order in error strength $\gamma$ (linear $\gamma$ term vanishes), proving the protocol to be robust against oscillator dephasing for $\beta$ values close to $0$ and around the thresholds with small deviations outside the threshold regions.

\subsection{Binary Decision for Quadratic Gaussian Signals} \label{sec:mp_ta-beta-squeezing}

If $\beta$ appears in the form of a squeezing-like signal rather than a simple displacement as discussed in the previous subsections, the gaussian signal can be obtained by setting $k=2$ for $S_{\beta}$ in Eq.~\eqref{s-beta}.
The $P(M = \downarrow| \beta)$ for such a scenario can be effectively expressed as,
\begin{align}
    P(M = \downarrow| \beta) 
    &=  
    \sum_{s= -d}^{d}  \sum_{l= -2d}^{2d} c_{s,l}
    e^{-4s^2\kappa^2\beta^2}
    e^{i(4\kappa^2\beta)ls}, 
    \\
     \nonumber
    \text{where,~~} 
   c_{s,l} &= \left[ \sum_{\substack{r,m=-d\\ r+m=l}}^{d} c_{s,r,m}
    \right],
    \\
    \text{and,~~}
    c_{s,r,m} 
    &=  \nonumber
      \sum_{n = -d}^{d}
    \left( p_n p^{*}_{m} +  q_n q^{*}_{m} \right)
        \\
    &\times \nonumber
    \left( p^{*}_{n + 2s} p_{m + 2r} +  q^{*}_{n + 2s} q_{m + 2r} \right) 
    e^{-(r-s)^2\kappa^2}.
\end{align}
Compared to the case of the displacement signal as seen in Eq.~\eqref{eq:csbeta} and \eqref{eq:qspi_eq_20}, two major differences arise. First, an additional Gaussian damping term $e^{-4s^2\kappa^2\beta^2}$ appears, which introduces a sense of localization over the parameter $\beta$. Secondly, the underlying phase variable changes from $e^{i(2\kappa\beta)s}$ to $e^{i(4\kappa^2\beta)ls}$, which introduces a dependency on two indices $s$ and $l$ rather than a single index $s$ in case of displacement signals.
This representation can be interpreted from the perspective of frame \cite{1d493e2c-cfa9-3224-9c3d-d3d9555183b1, 6723d674-1f20-39d1-a0fa-adb4f9f0358a} with family of functions 
\begin{align}
     g_{s,l}(\beta) &= \label{eq:frame-unit}
     e^{i(4\kappa^2\beta)ls} 
     e^{-(4s^2\kappa^2)\beta^2}.
\end{align}
Each function $g_{s,l}(\beta)$ consists of a Gaussian window $e^{-(4s^2\kappa^2)\beta^2}$ modulated by a complex exponential $e^{i(4\kappa^2\beta)ls}$, which is analogous to Gabor functions \cite{1573950400118324608, janssen1981gabor, bastiaans1980gabor}. Detailed discussion on frame analysis \cite{daubechies2002wavelet} using the obtained family of functions is presented in Appendix \ref{appndx:squeezed-signals-frame}.

However, unlike standard Gabor frames, where the Gaussian window has fixed width, the width here depends on the factor $s^2$. As $|s|$ increases, the Gaussian becomes more localized, indicating varying levels of resolution of the Gaussian kernels. This dependence introduces a multi-resolution behavior, that can allow different components of the expansion of $ P(M = \downarrow| \beta) $ to probe $\beta$ at different scales, thereby enhancing flexibility in representing and detecting structured signal features.

\section{Numerical Simulation Results} \label{sec:results}
\begin{figure}[ht]
    \centering
    \includegraphics[width=1\linewidth]{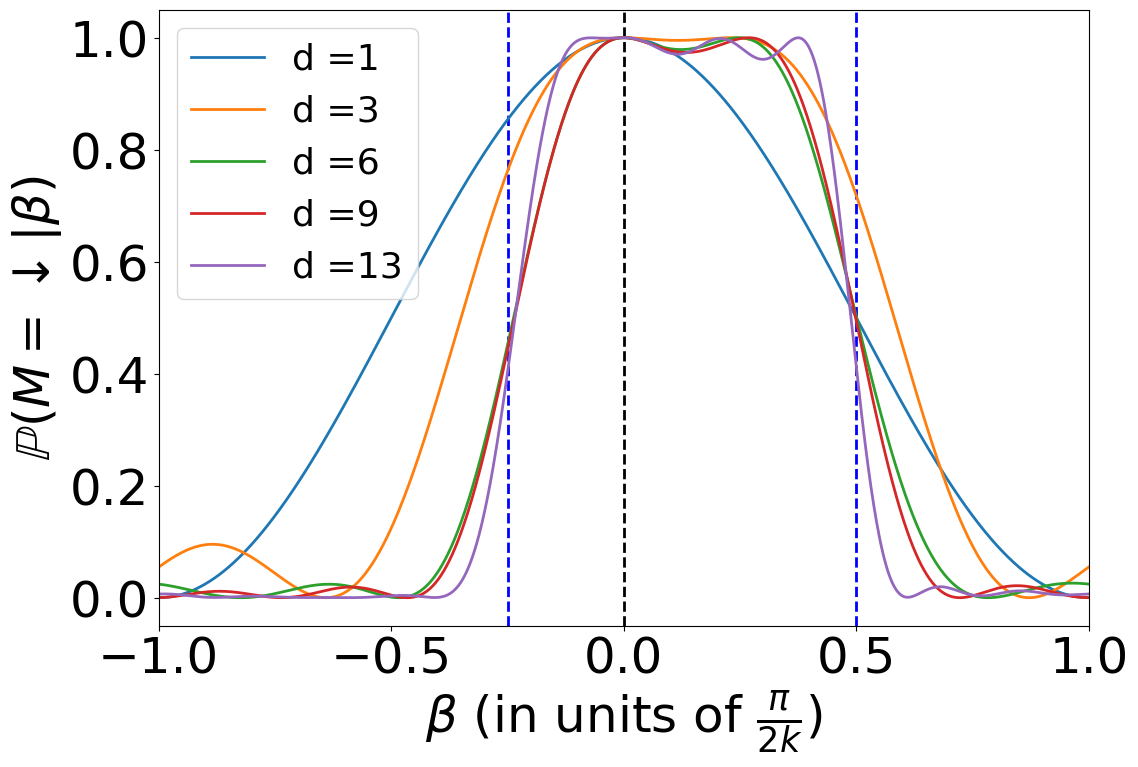}
     \vspace*{-0.5\baselineskip}
    \caption{Qubit repsonse function for asymmetric thresholding problem. $P(M = \downarrow | \beta)$ plotted against $\beta$ for $\kappa = \frac{1}{2048}$ for degrees $1,3, 6, 9, \text{and }13$ with the blue dotted lines representing the thresholds at $\beta_{-th} = -\frac{\pi}{8\kappa}$ and $\beta_{+th} = \frac{\pi}{4\kappa}$.}
    \label{fig:GQSPI-Sim-Results}
\end{figure}
\begin{figure}[ht]
    \centering
\includegraphics[width=1\linewidth]{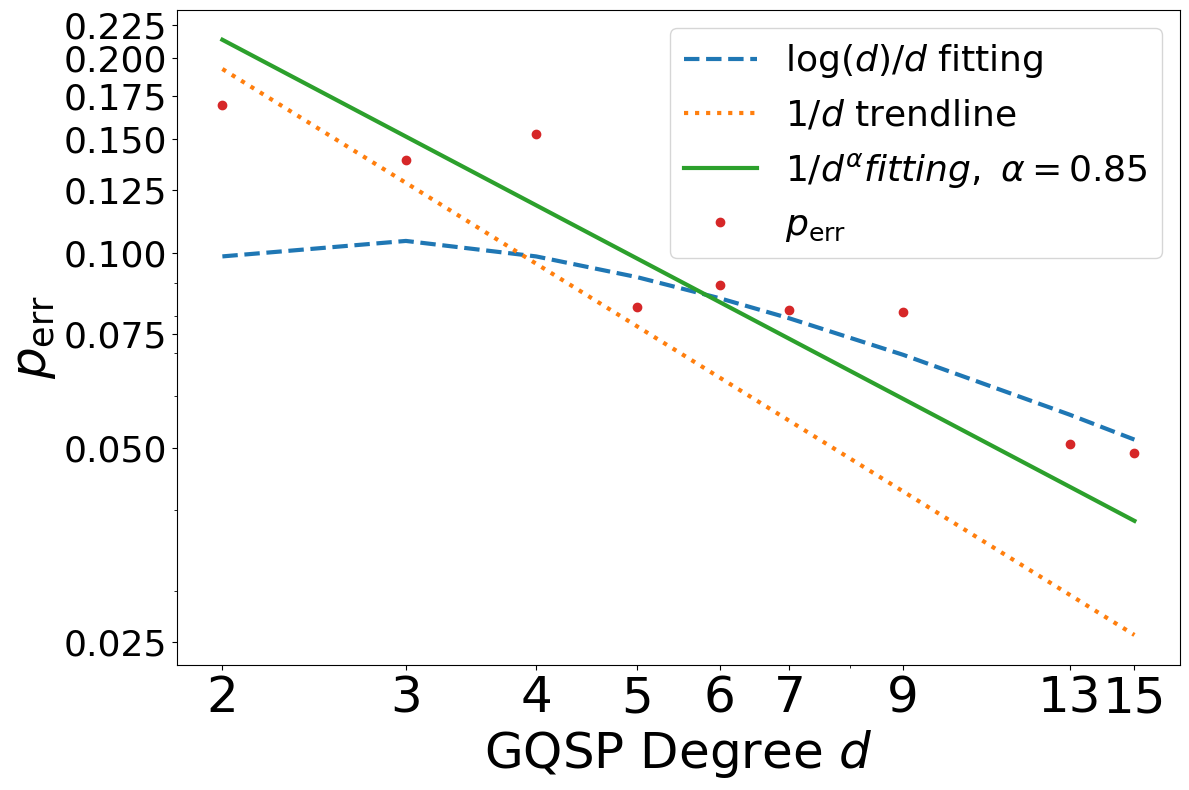}
    \vspace*{-0.5\baselineskip}
    \caption{Total probability of decision error $p_{\text{err}}$ vs GSQPI degree obtained from loss calculated for simulation results in Fig. \ref{fig:GQSPI-Sim-Results}. The red dots on log-log plot indicate the $p_{\text{err}}$ value for the given GQSP degree. The trendlines for $\frac{1}{d}, \frac{1}{d^\alpha}$ and fitting line for $\frac{1}{d}\log{d}$ are plotted against the given GQSP degrees to show the general trend of non-linear error suppression with increasing degree.}
    \label{fig:perr_vs_deg}
\end{figure}

We simulate the probability response function $P(M=\downarrow \mid \beta)$ for different GQSPI degrees with asymmetric thresholds, where we wish to detect if $\beta \in [-\frac{\pi}{8\kappa}, \frac{\pi}{4\kappa}]$, or outside of it with a small decision error. For a given degree, the phase parameters $\{\vec{\theta},\vec{\phi},\lambda_0\}$ are optimized using the Nelder-Mead optimizer to minimize the decision error $p_{\mathrm{err}}$ defined in Eq.~\eqref{eq:P-err-form}, until the desired error tolerance ($10^{-5}$ for this simulation) is achieved. From the results shown in Fig. \ref{fig:GQSPI-Sim-Results}, it can be observed that $P(M = \downarrow| \beta)$ is asymmetric with respect to $\beta = 0$ for GQSPI degrees greater than $1$ as expected. For degree $d=1$, the protocol reduces to a symmetric polynomial. The transition becomes sharper with increasing polynomial degree, and the total decision error decreases with increasing GQSP degree, as seen in Fig. \ref{fig:perr_vs_deg}, roughly following the expected behavior derived in Eq. \eqref{eq:degree-perr}.

\begin{figure}[ht]
    \centering
    \includegraphics[width=1\linewidth]{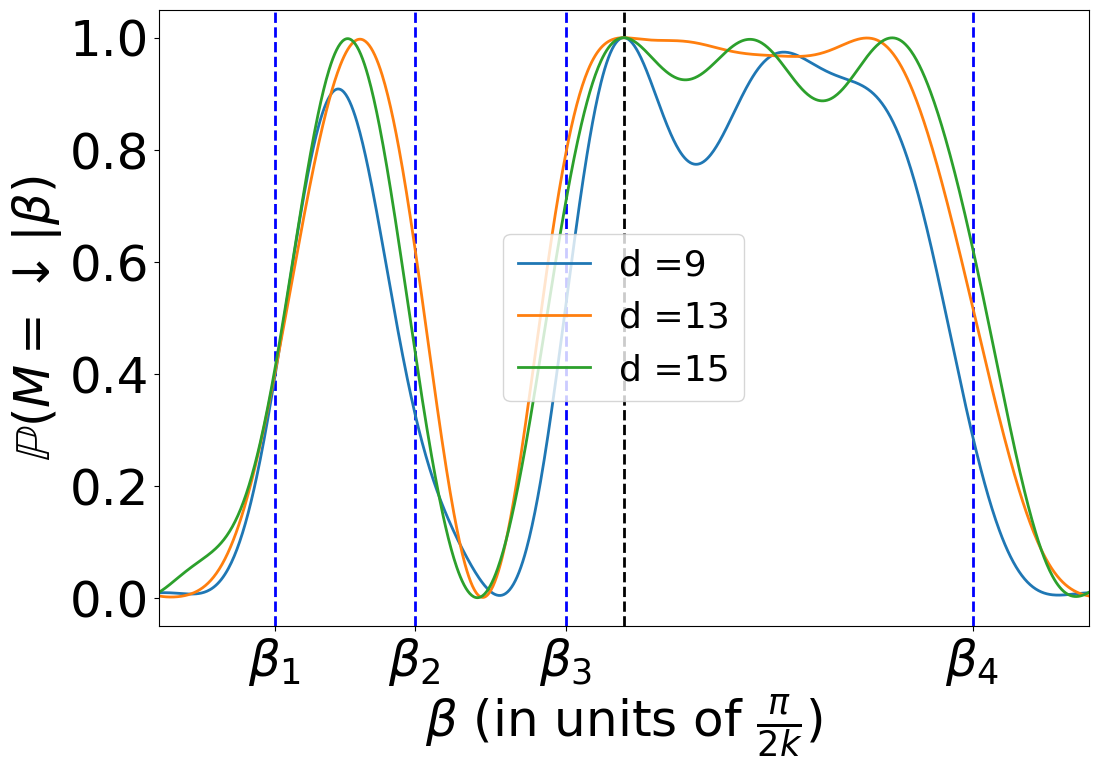}
    \vspace*{-0.8\baselineskip}
    \caption{Qubit response function for the multi-thresholding problem. $P(M = \downarrow | \beta)$ plotted against $\beta$ for $\kappa = \frac{1}{2048}$ for degrees $9, 13, \text{ and }15$ with the blue dotted lines representing the thresholds at $\beta_{1} = -0.75\frac{\pi}{2\kappa}$, $\beta_{2} = -0.45\frac{\pi}{2\kappa}, \beta_{3} = -0.125\frac{\pi}{2\kappa}$, $\beta_{4} = 0.75\frac{\pi}{2\kappa}$.
    The region for which we wish to detect $\beta$ is $[\beta_{1},\beta_{2}] \cup [\beta_{3},\beta_{4}]$.
    }
    \label{fig:double-window}
\end{figure}

In certain practical settings, the goal is to determine whether the Gaussian signal parameter $\beta$ belongs to one of several prescribed bands of values. This gives rise to a multi-threshold detection problem. Assuming a simple case of two non-overlapping ranges of thresholds, say $[\beta_{1},\beta_{2}]$, and $[\beta_{3},\beta_{4}]$, simulation results obtained in Fig. \ref{fig:double-window} show that GQSPI can be used for multi-thresholding problems as well.
Here the objective is to distinguish whether the displacement parameter $\beta$ lies inside either of these two bands or outside both of them. Multi-threshold detection generally requires higher GQSP degrees as compared to single-band detection. This is because each additional threshold introduces a transition region that must be accommodated in the polynomial approximation. When the desired bands are narrow or closely spaced, a higher-degree polynomial is required to capture the response while adhering to the intended error bounds in both the passband and stopband regions. Code for the simulation results obtained in this section can be found in Ref.~\cite{code_repo_gqspi}.

\section{Conclusions} \label{sec:disc}
The ability to actively detect a general quantum Gaussian signal over a single or multiple threshold bands in the presence of classical and quantum noise is a challenging practical problem. We address these aspects of quantum detection problem using the proposed GQSPI framework which is capable of systematically improving the decision accuracy while being robust against quantum dephasing noise. The key techniques that enabled this is to view the Gaussian signal detection problem from a polynomial approximation perspective under various frame and basis in phase space.

Several directions for future works include extending this framework to coupled multi-qubit, multi-oscillator systems which can be used to potentially interrogate non-commuting properties of a general signal in noisy settings. It also remains to be explored how this protocol integrates with practical quantum optical communication channels where detection of squeezed lights is of fundamental importance. The protocol can be further extended to incorporate higher-dimensional generalizations of QSP and GQSP \cite{rossi2022multivariable, laneve2024quantumsignalprocessingsun, lu2026quantum, 10.1063/5.0312254} to accommodate decision problems over multivariate Gaussian signals simultaneously with distinct thresholds. We look forward to these endeavors to uncover deeper connections between quantum detection theory and quantum algorithms, thereby open an exciting avenue for solving decision problems in the quantum domain.

\section*{Acknowledgment}
This work is supported by the U.S. Department of Energy, Office of Science, Advanced Scientific Computing Research, under contract number DE-SC0025384.



\balance
\bibliographystyle{ieeetr}
\bibliography{ref_isit}

\onecolumn
\appendices

\renewcommand{\thesubsection}{\thesection-\Roman{subsection}}
\makeatletter
\def\@seccntformat#1{%
  \ifcsname the#1\endcsname
    \csname the#1\endcsname\hskip 0.75em
  \fi}
\makeatother

\section{Generalized Quantum Signal Processing for Hybrid Qubit-Oscillator Systems} \label{sec:appndx-GQSP-Poly-Forms}

\subsection{Nature of Polynomials}

Assume  $\hat{\omega} = e^{i\kappa \hat{x}}$ for simplicity of calculation.
\\
For GQSP of degree 1, the polynomials obtained by simply multiplying one iteration of the qubit rotation gates and the conditional displacement gate results in the following polynomials,
\begin{align}
    P_1(\hat{x}) &= \label{eq:P1}
    e^{i(\phi_{1}+ \lambda_0)} \left(
        e^{i \phi_0} \cos{\theta_{1}} \cos{\theta_0} e^{i \kappa \hat{x}}
        +
       \sin{\theta_{1}}\sin{\theta_0} e^{-i \kappa \hat{x}}
        \right) 
        \\
        Q_1(\hat{x}) &= \label{eq:Q1}
    e^{i\lambda_0} \left(
        e^{i\phi_0} \sin{\theta_{1}}   \cos{\theta_0} e^{i \kappa \hat{x}}
        -
        \cos{\theta_{1}}\sin{\theta_0} e^{-i \kappa \hat{x}}
        \right).
\end{align}

It can be observed that the coefficients of the polynomial can be complex or real, depending on the choice of $\vec{\theta} = \{\theta_0, \theta_1 \cdots \theta_d\}, ~\vec{\phi} = \{\phi_0, \phi_1 \cdots \phi_d\}$, and $\lambda_0$.

Going from degree $d$ to $d+1$ involves multiplying with the following ``\textbf{unit}":
\begin{align}
    (R(\theta_i, \phi_i, 0)\otimes I_{osc})D_c(i\kappa/\sqrt{2}) 
    &= 
    \label{eq:fwd-unit}
    \begin{bmatrix} 
        e^{i \phi_{d+1}}\cos{\theta_{d+1}}e^{i \kappa \hat{x}} & e^{i\phi_{d+1}}\sin{\theta_{d+1}}e^{-i \kappa \hat{x}} \\
        \sin{\theta_{d+1}}e^{i \kappa \hat{x}} & 
        -\cos{\theta_{d+1}}e^{-i \kappa \hat{x}}
    \end{bmatrix}.
\end{align}

Thus the recursive relationship between the polynomials are: 
\begin{align}
    P_{d+1}(\hat{x}) 
    &= \label{eq:Pd-recurr}
        e^{i \phi_{d+1}}\cos{\theta_{d+1}}e^{i \kappa \hat{x}}P_d(\hat{x}) 
        + 
        e^{i\phi_{d+1}}\sin{\theta_{d+1}}e^{-i \kappa \hat{x}} Q_d(\hat{x})
        \\
    Q_{d+1}(\hat{x}) 
    &= \label{eq:Qd-recurr}
        \sin{\theta_{d+1}}e^{i \kappa \hat{x}}P_d(\hat{x}) -
        \cos{\theta_{d+1}}e^{-i \kappa \hat{x}}Q_d(\hat{x}) .
\end{align}

Combining the Eq. \eqref{eq:Pd-recurr}, \eqref{eq:Qd-recurr} and Eq. \eqref{eq:P1}, \eqref{eq:Q1}, and calculating the polynomial $P$ for degree 2 and 3,
\begin{align*}
    P_{1}(\hat{x}) &= p_1 e^{i \kappa \hat{x}} + p_{-1} e^{-i \kappa \hat{x}} 
    \\
    P_{2}(\hat{x}) &= p_{-2}e^{-i 2 \kappa \hat{x}} + p_{2}e^{i 2 \kappa \hat{x}} + p_0
    \\
    P_{3}(\hat{x}) &= p_{3} e^{i 3\kappa \hat{x}} + p_{-3} e^{-i 3\kappa \hat{x}} + p_{-1} e^{-i \kappa \hat{x}} + p_{1} e^{i \kappa \hat{x}}.
\end{align*}

Thus, the general form of the polynomial is 
\begin{align}
    P_d(\hat{x}) = \sum_{n=-d}^{d} p_n \hat{x}^n, \quad n = \{-d, -d+2, \cdots, d-2, d\}.
    \label{eq:P-general}
\end{align}
This polynomial form is obtained for any operator $\hat\omega$.

\subsection{Recursive Relationship Between the GQSP Coefficients}

Expanding both sides of Eq. \eqref{eq:Pd-recurr} using the general form of polynomial given in Eq \eqref{eq:P-general}, 
\begin{align*}
    \sum_{r=-(d+1)}^{d+1} p^{(d+1)}_r e^{i r \kappa \hat{x}}
    &= 
    e^{i \phi_{d+1}}\cos{\theta_{d+1}}e^{i \kappa \hat{x}}
    \left(\sum_{r=-(d)}^{d} p^{(d)}_re^{i r \kappa \hat{x}}
    \right)
    + e^{i\phi_{d+1}}\sin{\theta_{d+1}}e^{-i \kappa \hat{x}} 
        \left(\sum_{r=-(d)}^{d} q^{(d)}_re^{i r \kappa \hat{x}}
        \right)
    \\
    &=
    e^{i \phi_{d+1}}\cos{\theta_{d+1}}
    \left( p^{(d)}_de^{i (d+1) \kappa \hat{x}} + p^{(d)}_{d-2}e^{i (d-1) \kappa \hat{x}} +
    \cdots
    +
     p^{(d)}_{-d}e^{i (-d+1) \kappa \hat{x}}
    \right)
    \\
    &+
    e^{i\phi_{d+1}}\sin{\theta_{d+1}}
    \left(
     q^{(d)}_de^{i (d-1) \kappa \hat{x}} + q^{(d)}_{d-2}e^{i (d-3) \kappa \hat{x}} +
    \cdots
    +
     q^{(d)}_{-d}e^{-i (d+1) \kappa \hat{x}}
    \right)
\end{align*}
Comparing coefficients on both sides results in the following relations,
\begin{align*}
    p^{(d+1)}_r = 
    \begin{cases}
        e^{i\phi_{d+1}} \cos{\theta_{d+1}}p^{(d)}_{r-1}, 
        \quad r = d+1
        \\
        e^{i\phi_{d+1}} \sin{\theta_{d+1}}q^{(d)}_{r+1},
        \quad r = -(d+1)
        \\
        e^{i\phi_{d+1}}\cos{\theta_{d+1}}p^{(d)}_{r-1}
        + 
        e^{i\phi_{d+1}}\sin{\theta_{d+1}}q^{(d)}_{r+1}, \quad |r| \le (d-1)
    \end{cases},
\end{align*}
Using a similar approach for coefficients of the polynomial $Q$, 
\begin{align*}
    q^{(d+1)}_r = 
    \begin{cases}
        \sin{\theta_{d+1}}p^{(d)}_{r-1}, 
        \quad r = d+1
        \\
        -\cos{\theta_{d+1}}q^{(d)}_{r+1},
        \quad r = -(d+1)
        \\
        \sin{\theta_{d+1}}p^{(d)}_{r-1}
        -
        \cos{\theta_{d+1}}q^{(d)}_{r+1}, \quad |r| \le (d-1)
    \end{cases}.
\end{align*}

\section{Generalized Quantum Signal Processing Interferometry Protocol} \label{sec:appndx-GQSPI}

\subsection{Structure of \texorpdfstring{$U(\beta, \kappa)$}{}}

$U(\beta, \kappa)$ is the overall unitary obtained in the protocol by sandwiching the displacement signal $S_{\beta}$ by a $d$-degree GQSPI and its inverse,
\begin{align}
     U(\beta, \kappa) \nonumber
     &= \nonumber G^{-1}_{d,\vec{\theta}, \vec{\phi}, \lambda_0}(\omega) ~S_{\beta}~ G_{d,\vec{\theta}, \vec{\phi}, \lambda_0}(\omega)\\
     &= \nonumber
     \begin{bmatrix}
        \sum_{n=-(d)}^{d} p_n^{*} e^{-i n\kappa \hat{x}} 
        &
         \sum_{n=-(d)}^{d} q_n^{*} e^{-i n\kappa \hat{x}}
         \\
        -\sum_{n=-(d)}^{d} q_n e^{i n\kappa \hat{x}}
        &
         \sum_{n=-(d)}^{d} p_n e^{i n\kappa \hat{x}}
    \end{bmatrix}
    \begin{bmatrix}
        e^{i \beta \hat{p}}  & 0 \\
        0 & e^{i \beta \hat{p}}
    \end{bmatrix}
    \begin{bmatrix}
        \sum_{n=-(d)}^{d} p_n e^{i n\kappa \hat{x}} 
        &
         -\sum_{n=-(d)}^{d} q^{*}_n e^{-i n\kappa \hat{x}}
         \\
        \sum_{n=-(d)}^{d} q_n e^{i n\kappa \hat{x}}
        &
         \sum_{n=-(d)}^{d} p^{*}_n e^{-i n\kappa \hat{x}}
    \end{bmatrix}
     \\
     &= \label{eq:U-beta}
     \sum_{n,m=-d}^{d} C_{nm} e^{- i n \kappa \hat{x}} S_{\beta} e^{i m \kappa \hat{x}}, \quad \text{where }
     C_{nm} = 
    \begin{bmatrix}
        p^{*}_n p_m +  q^{*}_n q_m
        &
        -  \left(
        p^{*}_n q^{*}_m 
        - q^{*}_n p^{*}_m
        \right)
        \\
        \left( 
        p_n q_m
        -
        q_n p_m 
        \right)
        &
        q_n q^{*}_m +  p_n p^{*}_m
    \end{bmatrix}
\end{align}
where $n,m = \{-d, -d+2, \cdots, d-2, d\}$.

The top-left element of the operator $U(\beta, \kappa)$ can be expressed as,

\begin{align}
    U(\beta, \kappa)_{00} &= \label{eq:u00}
    \sum_{n,m=-d}^{d} \left( p^{*}_n p_m +  q^{*}_n q_m \right)e^{-i n\kappa \hat{x}} e^{i\beta \hat{p}} e^{i m \kappa \hat{x}}
    \\
    &= \nonumber
    \sum_{n,m=-d}^{d} \left( p^{*}_n p_m +  q^{*}_n q_m \right)
    e^{i\beta \hat{p}} e^{-i n\kappa \hat{x}} e^{-i n\kappa \beta}  e^{i m \kappa \hat{x}}
\end{align}

\subsection{Qubit Probability Response Function \texorpdfstring{$P(M = \downarrow| \beta)$}{}} 
\label{sec:prob-beta}

Using the equation obtained in Eq.~\eqref{eq:u00}, the qubit probability response function $P(M = \downarrow| \beta)$ can be expressed as,
\begin{align}
    P(M = \downarrow| \beta) 
    &= \nonumber
   \langle 0|_{osc} U^{\dagger}(\beta, \kappa)_{00}U(\beta, \kappa)_{00}|0 \rangle _{osc}
   \\
   &= \nonumber
   \int_{-\infty}^{\infty} 
   \left[
   \sum_{n, m, n', m' = -d}^{d} 
    \left( p_n p^{*}_m +  q_n q^{*}_m \right)\left( p^{*}_{n'} p_{m'} +  q^{*}_{n'} q_{m'} \right) e^{-i(n - n')\kappa \beta}
    e^{i (n - n' - m + m') \kappa \hat{x}}  \right]
   \psi^{2}_0(x) dx
   \\
  &=  \nonumber
  \sum_{n, m, n', m' = -d}^{d} 
  \left[
    \left( p_n p^{*}_m +  q_n q^{*}_m \right)\left( p^{*}_{n'} p_{m'} +  q^{*}_{n'} q_{m'} \right) 
      e^{-i(n - n')\kappa \beta}
      \int_{-\infty}^{\infty}
    e^{i (n - n' - m + m') \kappa x} 
   \psi^{2}_0(x) dx 
   \right] \\
   &= \nonumber
   \sum_{n, m, n', m' = -d}^{d} 
  \left[
    \left( p_n p^{*}_m +  q_n q^{*}_m \right)\left( p^{*}_{n'} p_{m'} +  q^{*}_{n'} q_{m'} \right) 
     e^{-\frac{1}{4} k^2 (n - n' - m + m')^2} \right] e^{-i(n - n')\kappa \beta} 
     \\
     &= \nonumber
      \sum_{s = -d}^{d} 
      \left(
      \sum_{n, m, r = -d}^{d}
     \left[
    \left( p_n p^{*}_{m} +  q_n q^{*}_{m} \right)\left( p^{*}_{n + 2s} p_{m + 2r} +  q^{*}_{n + 2s} q_{m + 2r} \right) 
     e^{- k^2 (r - s )^2} \right] 
      \right)e^{i(2\kappa \beta) s}
      \\
      &= 
      \sum_{s = -d}^{d} c_s \nu^s(\beta), 
      \quad \text{where }
       c_s =
      \sum_{n, m, r = -d}^{d}
    \left( p_n p^{*}_{m} +  q_n q^{*}_{m} \right)\left( p^{*}_{n + 2s} p_{m + 2r} +  q^{*}_{n + 2s} q_{m + 2r} \right) 
     e^{- k^2 (r - s)^2}
\end{align}

Here we have performed the variable substitution $n' - n = 2s$ and $m' - m = 2r$  with  $s, r \in [-d, d]$. Thus, $n' \rightarrow n + 2s$, and $m' \rightarrow 2r + m$.

We use the result $\int_{-\infty}^{\infty} e^{bx-ax^2} dx= e^{-b^2/4a}\sqrt{\frac{\pi}{a}} 
\rightarrow 
\int_{-\infty}^{\infty} e^{-i(m + n' - m' - n) \kappa x  -x^2}/\sqrt{\pi}dx   = e^{\frac{[-i(m + n' - m' - n) \kappa]^2}{4}} $.

$\psi_0(x) = \pi^{-1/4}e^{-x^2/2}$ is the vacuum state of the oscillator $\Rightarrow \psi^2_0(x) = e^{-x^2}/\sqrt{\pi}$.

\subsection{\texorpdfstring{$P_{ideal}$}{} and Decision Error Density \texorpdfstring{$p_{err}(\beta_{-th}, \beta_{+th}, \kappa)$}{} for Asymmetric \texorpdfstring{$\beta$}{}} 
\label{sec:prob-err-density}

The ideal qubit probability response function as given in Eq.~\eqref{eq:p_ideal},
\begin{align}
    P_{ideal} = \begin{cases}
    1 , \quad \beta_{-th} \leq \beta \leq \beta_{+th}
    \\
    0, \quad \text{otherwise}
\end{cases}
\end{align}

The decision error density $p_{err}(\beta_{-th}, \beta_{+th}, \kappa)$ would be the average of the difference between $P_{ideal}$ and the polynomial approximation obtained from the GQSPI protocol $P(M = \downarrow | \beta)$. The Eq. \eqref{eq:P-err-form} can be obtained as follows:

\begin{align}
    &p_{err}(\beta_{-th}, \beta_{+th}, \kappa) 
    = \nonumber
    \frac{\kappa}{\pi} 
    \int_{-\frac{\pi}{2\kappa}}^{\frac{\pi}{2\kappa}} 
    |P_{ideal} - P(M = \downarrow | \beta)|d\beta
    \\
    &= 
    \frac{\kappa}{\pi} 
    \left[
    \int_{-\frac{\pi}{2\kappa}}^{\beta_{-th}} P(M = \downarrow | \beta) d\beta
    +
     \int_{\beta_{-th}}^{\beta_{+th}} |1 - P(M = \downarrow | \beta)|d\beta
     +
      \int_{\beta_{+th}}^{\frac{\pi}{2\kappa}} P(M = \downarrow | \beta) d\beta
    \right]
    \\
    &= \nonumber
    \frac{\kappa}{\pi} 
    \left[
    \int_{-\frac{\pi}{2\kappa}}^{\beta_{-th}} 
    \left(
     \sum_{s = -d}^{d} c_s e^{i(2\kappa \beta) s}
    \right) d\beta
    +
     \int_{\beta_{-th}}^{\beta_{+th}} \left[
     1 - \left(
     \sum_{s = -d}^{d} c_s e^{i(2\kappa \beta) s}
    \right)
    \right]
    d\beta
     +
      \int_{\beta_{+th}}^{\frac{\pi}{2\kappa}}\left(
     \sum_{s = -d}^{d} c_s e^{i(2\kappa \beta) s}
    \right) d\beta
    \right]
    \\
    &=
    \nonumber
    \frac{ (\beta_{+th} - \beta_{-th})\kappa}{\pi}
    +
    \sum_{s = -d}^{d} c_s \text{sinc}(\pi s)
    +
     i \sum_{s = -d}^{d} c_s 
     \frac{e^{i( 2\kappa \beta_{+th}) s} - e^{i( 2\kappa \beta_{-th}) s}}{\pi s}
     \\
     &= \nonumber
      \frac{ (\beta_{+th} - \beta_{-th})\kappa}{\pi}
    +
    \sum_{s = -d}^{d} c_s H_s, 
    \quad \quad \text{where}~H_s = \text{sinc}(\pi s) + i \frac{e^{i( 2\kappa \beta_{+th}) s} - e^{i( 2\kappa \beta_{-th}) s}}{\pi s}
\end{align}

\subsection{Multiple Threshold Problem} \label{sec:multi-threshold}

In the example case of detecting $\beta$ over two non-overlapping ranges of thresholds, say $[\beta_{1},\beta_{2}]$, $[\beta_{3},\beta_{4}]$, then, only changes are in the $p_{err}$ calculation. The ideal function in this scenario is,
\begin{align}
    P_{ideal} = \begin{cases}
    1 , \quad \beta_{1} \leq \beta \leq \beta_{2} \cup \beta_{3} \leq \beta \leq \beta_{4}
    \\
    0, \quad \text{otherwise}
\end{cases}
\end{align}
\begin{figure}[h]
    \centering
    \includegraphics[scale=0.6]{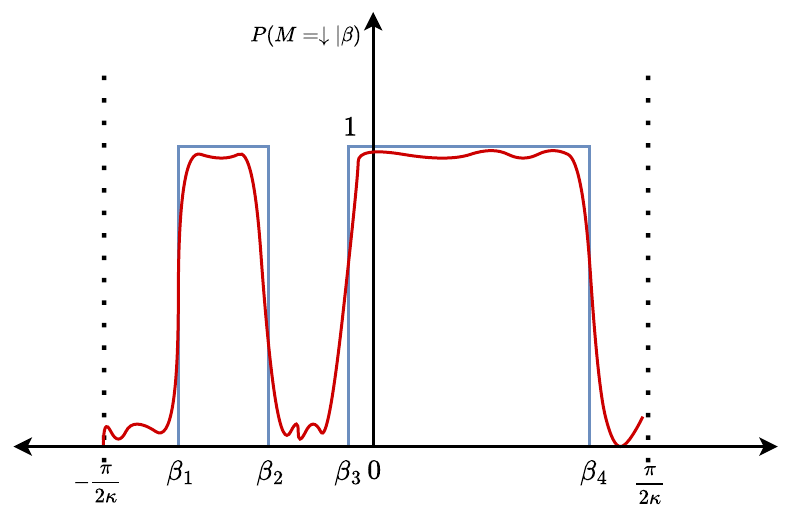}
    \caption{Example case of two bands of thresholds}
    \label{fig:gqspi-response-double}
\end{figure}
Taking a similar approach as shown in the previous Sec.~\ref{sec:prob-err-density},
\begin{align}
    &p_{err}(\beta_{1}, \beta_{2}, \beta_{3}, \beta_{4}, \kappa) 
    = \nonumber
    \frac{\kappa}{\pi} 
    \int_{-\frac{\pi}{2\kappa}}^{\frac{\pi}{2\kappa}} 
    |P_{ideal} - P_{approx}(\beta)|d\beta
    \\
    &= \nonumber
    \frac{\kappa}{\pi} 
    \left[
    \int_{-\frac{\pi}{2\kappa}}^{\beta_{1}} P(M = \downarrow| \beta) d\beta
    +
     \int_{\beta_{1}}^{\beta_{2}} |1 - P(M = \downarrow| \beta)|d\beta
     +
      \int_{\beta_{2}}^{\beta_{3}} P(M = \downarrow| \beta) d\beta
      +
     \int_{\beta_{3}}^{\beta_{4}} |1 - P(M = \downarrow| \beta)|d\beta
     +
     \int_{\beta_{4}}^{\frac{\pi}{2\kappa}} P(M = \downarrow| \beta) d\beta
    \right]
    \\
    &= \nonumber
     \frac{\kappa}{\pi} 
    \left[
    \int_{-\frac{\pi}{2\kappa}}^{\beta_{1}} \left(
     \sum_{s = -d}^{d} c_s e^{i(2\kappa \beta) s}
    \right) d\beta
    +
     \int_{\beta_{1}}^{\beta_{2}} |1 - \left(
     \sum_{s = -d}^{d} c_s e^{i(2\kappa \beta) s}
    \right)|d\beta
    +
      \int_{\beta_{2}}^{\beta_{3}} \left(
     \sum_{s = -d}^{d} c_s e^{i(2\kappa \beta) s}
    \right) d\beta
    \right.
    \\ \nonumber
    &\left.
      +
     \int_{\beta_{3}}^{\beta_{4}} |1 - \left(
     \sum_{s = -d}^{d} c_s e^{i(2\kappa \beta) s}
    \right)|d\beta
     +
     \int_{\beta_{4}}^{\frac{\pi}{2\kappa}} \left(
     \sum_{s = -d}^{d} c_s e^{i(2\kappa \beta) s}
    \right) d\beta
    \right]
     \\
     &= \label{eq:multi-thesh-p_err}
     \frac{\kappa}{\pi} (\beta_{4} - \beta_{3} + \beta_{2} - \beta_{1})
     +
     \sum_{s = -d}^{d} c_s 
     \left[
     \text{sinc}(\pi s)
     +
     (\frac{i}{\pi s}) (e^{i( 2\kappa \beta_{2}) s} - 
     e^{i( 2\kappa \beta_{1}) s} 
     +
     e^{i( 2\kappa \beta_{4}) s} - 
     e^{i( 2\kappa \beta_{3}) s} )
     \right]
\end{align}

This form of $p_{err}(\beta_{1}, \beta_{2}, \beta_{3}, \beta_{4}, \kappa)$ obtained in Eq.~\eqref{eq:multi-thesh-p_err} can also be generalized for more than 2 bands of thresholds. 
Assume that $\beta$ is required to be detected for $m$ number of non-overlapping bands of thresholds indexed by pairs $\{\vec{\beta}_{-th}, \vec{\beta}_{+th}\} = \{ \{\beta_{1,-th}, \beta_{1,+th}\}, \{\beta_{2,-th}, \beta_{2,+th}\}, \cdots, \{\beta_{m,-th}, \beta_{m,+th}\}\}$, then the expression for $p_{err}$ becomes,
\begin{align*}
    p_{err}(\{\vec{\beta}_{-th}, \vec{\beta}_{+th}\}, \kappa) 
    = 
    \frac{\kappa}{\pi} \sum_{j=1}^{m}(\beta_{j,+th} - \beta_{j,-th})
    +
    \sum_{s = -d}^{d} c_s 
     \left[
     \text{sinc}(\pi s)
     +
     (\frac{i}{\pi s}) \sum_{j=1}^{m} \left(
     e^{i( 2\kappa \beta_{j,+th}) s} - 
     e^{i( 2\kappa \beta_{j,-th}) s}\right)
     \right]
\end{align*}
as shown previously in Eq.~\eqref{eq:general-multi-thresh-perr}.

\subsection{Symmetric \texorpdfstring{$ P(M = \downarrow| \beta)$ for degree 1}{}} \label{sec:gqspi-degree1-symmetry}
The operator sequence for the degree-1 GQSPI set-up would be given by $U(\beta, \kappa):$
\begin{align*}
    \left[(R(\theta_0, -\lambda_0, -\phi_0)\otimes I_{osc})D_c(-i\kappa/\sqrt{2})\right]
    S_{\beta}
    \left[D_c(i\kappa/\sqrt{2})(R(\theta_0, \phi_0, \lambda_0)\otimes I_{osc})\right]
\end{align*}

\begin{align*}
     U(\beta, \kappa) 
     &= 
     \begin{bmatrix} 
        e^{-i(\lambda + \phi)}\cos{\theta}e^{-i \kappa \hat{x}} 
        & e^{-i\lambda}\sin{\theta}e^{i \kappa \hat{x}} \\
        e^{-i\phi}\sin{\theta}e^{-i \kappa \hat{x}} 
        & 
        -\cos{\theta}e^{i \kappa \hat{x}}
    \end{bmatrix}
    \begin{bmatrix}
        e^{i \beta \hat{p}}e^{i \kappa \hat{x}}e^{i(\lambda+ \phi)}\cos{\theta} & e^{i \beta \hat{p}}e^{i \kappa \hat{x}}e^{i\phi}\sin{\theta} \\
        e^{i \beta \hat{p}}e^{-i \kappa \hat{x}} e^{i\lambda}\sin{\theta} & 
        -e^{i \beta \hat{p}}e^{-i \kappa \hat{x}}\cos{\theta}
    \end{bmatrix}
    \\
    &= 
    \begin{bmatrix} 
    e^{-i \kappa \hat{x}} e^{i \beta \hat{p}}e^{i \kappa \hat{x}}
    \cos^2{\theta} +
    \sin^2{\theta}e^{i \kappa \hat{x}}e^{i \beta \hat{p}}e^{-i \kappa \hat{x}}
    &
e^{-i\lambda }\cos{\theta}\sin{\theta}
[
e^{-i \kappa \hat{x}}e^{i \beta \hat{p}}e^{i \kappa \hat{x}}
- e^{i \kappa \hat{x}}e^{i \beta \hat{p}}e^{-i \kappa \hat{x}}
]
    \\
       e^{i\lambda}\cos{\theta}\sin{\theta}[ e^{-i \kappa \hat{x}}  e^{i \beta \hat{p}}e^{i \kappa \hat{x}}
        - e^{i \kappa \hat{x}} e^{i \beta \hat{p}}e^{-i \kappa \hat{x}} ]
         &
         \sin^2{\theta}e^{-i \kappa \hat{x}} e^{i \beta \hat{p}}e^{i \kappa \hat{x}}
        +
        \cos^2{\theta}e^{i \kappa \hat{x}}
        e^{i \beta \hat{p}}e^{-i \kappa \hat{x}}
    \end{bmatrix}\\
    &= \begin{bmatrix}
    (e^{i \kappa \beta} \cos^2{\theta}
     +
    e^{-i \kappa \beta}\sin^2{\theta})e^{i \beta \hat{p}}
    &
    e^{-i\lambda }\cos{\theta}\sin{\theta}
(e^{i \kappa \beta}- e^{-i \kappa \beta})e^{i \beta \hat{p}}
\\
 e^{i\lambda}\cos{\theta}\sin{\theta}( e^{i \kappa \beta}
        - e^{-i \kappa \beta} )e^{i \beta \hat{p}}
        &
        (\sin^2{\theta} e^{i \kappa \beta} 
        +
        \cos^2{\theta}e^{-i \kappa \beta})
        e^{i \beta \hat{p}}
    \end{bmatrix}  
\end{align*}
\textbf{Input state:} $\ket{\downarrow}\ket{0}_{osc}$
\\
\textbf{Output state:} $ U(\beta, \kappa) \ket{\downarrow} \ket{0}_{osc} = 
\begin{bmatrix}
e^{i \kappa \beta} \cos^2{\theta} + e^{-i \kappa \beta}\sin^2{\theta}
    \\
    e^{i\lambda}\cos{\theta}\sin{\theta}( e^{i \kappa \beta}
        - e^{-i \kappa \beta} )
\end{bmatrix} \otimes e^{i \beta \hat{p}}\ket{0}_{osc}$

Let $\nu(\beta) = e^{i 2\kappa \beta}$,
\begin{align*}
    Prob(M = \downarrow|\beta) 
    &= \langle 0|_{osc} U(\beta, \kappa)^{\dagger}_{00}U(\beta, \kappa)_{00} |0\rangle_{osc} \\
    &= (e^{-i \kappa \beta} \cos^2{\theta} + e^{i \kappa \beta}\sin^2{\theta})(e^{i \kappa \beta} \cos^2{\theta} + e^{-i \kappa \beta}\sin^2{\theta})
    \\
    &= \cos^4{\theta}+ e^{-i 2\kappa \beta}\cos^2{\theta}\sin^2{\theta}+ e^{i 2\kappa \beta}\cos^2{\theta}\sin^2{\theta} +  \sin^4{\theta}\\
    &= (\cos^4{\theta}+\sin^4{\theta}) + \cos^2{\theta}\sin^2{\theta}\nu^{-1}(\beta) + \cos^2{\theta}\sin^2{\theta}\nu(\beta)\\
    &=c_0 + c_{-1}\nu^{-1}(\beta) + c_{1}\nu(\beta)
\end{align*}
For degree 1, $c_{-1} = c_{1}$, the response function is symmetric about $\beta=0$ and only depends on $\theta_0$ value. This behavior is also confirmed by the simulation results in Fig. \ref{fig:GQSPI-Sim-Results}.

\section{GQSPI Protocol for Stochastic \texorpdfstring{$\beta$}{}}
\label{sec:appndx-gaussian-beta}
\subsection{Expression of \texorpdfstring{$p_{err}(\beta_{-th}, \beta_{+th}, \kappa, \sigma_{\beta}, \mu_{\beta})$}{}}

Let $\beta$ follow some Gaussian distribution with mean $\mu_{\beta}$ and variance $\sigma^2_{\beta}$ given by, 
\begin{align*}
    \beta \sim \frac{1}{\sqrt{2\pi \sigma^2_{\beta}}}e^{-\frac{(\beta - \mu_{\beta})^2}{2\sigma^2_{\beta}}}
\end{align*}

Then the average decision error probability density for the protocol would be, 

\begin{align}
   & \nonumber
   p_{err}(\beta_{-th}, \beta_{+th}, \kappa, \sigma_{\beta}, \mu_{\beta})
   = \frac{\kappa}{\pi}\mathbb{E} \left[ 
    |P_{ideal} - P_{approx}(\beta) |\right  ]
     \\
     &= \nonumber
     \frac{\kappa}{\pi}
     \sum_{s = -d}^{d}  \frac{c_s}{\sqrt{2\pi \sigma^2_{\beta}}} \left[
     \int_{-\frac{\pi}{2\kappa}}^{\beta_{-th}}
     e^{i(2\kappa \beta) s}
     e^{-\frac{(\beta - \mu_{\beta})^2}{2\sigma^2_{\beta}}} d\beta 
     +
     \int_{\beta_{+th}}^{\frac{\pi}{2\kappa}}
     e^{i(2\kappa \beta) s}
     e^{-\frac{(\beta - \mu_{\beta})^2}{2\sigma^2_{\beta}}} d\beta 
     \right]
     \\
      &+ \nonumber
       \frac{\kappa}{\pi}
      \int_{\beta_{-th}}^{\beta_{+th}}
     \left(1- \sum_{s = -d}^{d}  \frac{c_s}{\sqrt{2\pi \sigma^2_{\beta}}} e^{i(2\kappa \beta) s}
     e^{-\frac{(\beta - \mu_{\beta})^2}{2\sigma^2_{\beta}}}
     \right) d\beta 
       \\
       &= \nonumber
       \frac{\kappa}{\pi}
     (\beta_{+th} - \beta_{-th})
     - \frac{\kappa}{\pi}
     \sum_{s = -d}^{d}  c_s
     e^{-2\kappa^2 s^2 \sigma^2_{\beta}} 
     e^{i(2\kappa s \mu_{\beta})}
     \left[
      \text{erf}\left(\frac{\beta_{+th} - \mu_{\beta}}{\sqrt{2}\sigma_{\beta}} -i \sqrt{2} \kappa s \sigma_{\beta}\right)
      -
       \text{erf}\left(\frac{\beta_{-th} - \mu_{\beta}}{\sqrt{2}\sigma_{\beta}} -i \sqrt{2} \kappa s \sigma_{\beta}\right)
       \right]
       \\
       &+ \label{eq:perr-beta-stochastic}
       \frac{\kappa}{\pi}
     \sum_{s = -d}^{d}  \frac{c_s}{2} 
     e^{-2\kappa^2 s^2 \sigma^2_{\beta}} 
     e^{i(2\kappa s \mu_{\beta})}
     \left[
      \text{erf}\left(\frac{\frac{\pi}{2\kappa} - \mu_{\beta}}{\sqrt{2}\sigma_{\beta}} -i \sqrt{2} \kappa s \sigma_{\beta}\right)
      -
       \text{erf}\left(\frac{-\frac{\pi}{2\kappa} - \mu_{\beta}}{\sqrt{2}\sigma_{\beta}} -i \sqrt{2} \kappa s \sigma_{\beta}\right)
       \right]
\end{align}
As seen in the previous sections, the only tweak required is updating the probability of decision error density in order to find appropriate phase angles to accommodate prior information on $\beta$.

\section{GQSPI Protocol in Presence of Dephasing Noise on Oscillator}\label{sec:appnx-gqspi-deph}

\subsection{GQSP Form for Dephasing Error}
Assume that the a dephasing error occurs after the application of each conditional displacement on the oscillator, then the GQSPI protocol gets modified in the following manner,
    \begin{align*}
    G_d(\vec{\theta}, \vec{\phi}, \lambda_0, \vec{\gamma}) 
    &= 
    \left(\prod_{i=1}^{d} (R(\theta_i, \phi_i, 0)\otimes I_{osc}) \mathcal{D}_c(i \kappa / \sqrt{2}) R_{osc}(\gamma_i) \right)(R(\theta_0, \phi_0, \lambda_0)\otimes I_{osc})
    \\
    &= \nonumber
    (R(\theta_d, \phi_d, 0)\otimes I_{osc})\mathcal{D}_c(i \kappa / \sqrt{2}) R_{osc}(\gamma_d) (R(\theta_{d-1}, \phi_{d-1}, 0)\otimes I_{osc})\mathcal{D}_c(i \kappa / \sqrt{2}) R_{osc}(\gamma_{d-1}) 
    \\
    & \nonumber
    \cdots \mathcal{D}_c(i \kappa / \sqrt{2}) R_{osc}(\gamma_{1})(R(\theta_0, \phi_0, \lambda_0)\otimes I_{osc})
    \\
    &=
    R_{osc}(\sum_{i=1}^{d}\gamma_i)\left(
    \prod_{i=1}^{d}
    (R(\theta_i, \phi_i, 0)\otimes I_{osc})\mathcal{D}_c(\alpha'_i)
    \right) (R(\theta_0, \phi_0, \lambda_0)\otimes I_{osc})
        \\
    &\text{where,}\quad \alpha'_i = \frac{i\kappa \cos\Gamma_i + \kappa \sin\Gamma_i}{\sqrt{2}} 
    = \frac{\kappa}{\sqrt{2}}e^{i(\frac{\pi}{2} -\Gamma_i)}, 
    \quad \text{and,}\quad
    \Gamma_i = \sum_{j=1}^{i} \gamma_j.
    \end{align*}
The above form can be obtained by shifting the oscillator dephasing rotation gates $R_{osc}(\gamma)$ towards the left hand side.

\subsection{Polynomial Form of GQSP with Dephasing}

Polynomial obtained for degree 1 are of the form:
\begin{align*}
     P_1(\vec{\gamma}) &= 
     e^{i(\phi_{1}+ \lambda_0)} \left(
        e^{i \phi_0} \cos{\theta_{1}} \cos{\theta_0} \mathcal{D}(\alpha'_1)
        +
       \sin{\theta_{1}}\sin{\theta_0} \mathcal{D}(-\alpha'_1)
        \right)
        \\
    Q_1(\vec{\gamma}) &= e^{i\lambda_0} \left(
        e^{i\phi_0} \sin{\theta_{1}}   \cos{\theta_0} \mathcal{D}(\alpha'_1)
        -
        \cos{\theta_{1}}\sin{\theta_0} \mathcal{D}(-\alpha'_1)
        \right)
\end{align*}

Similarly, polynomial obtained for degree 2 are:
\begin{align*}
    P_2(\vec{\gamma})  &= p_{2}
     p_{2}
     \mathcal{D}(\alpha'_{2} + \alpha'_{1})
      e^{-i\frac{\kappa^2}{2}\sin(\Gamma_2 - \Gamma_1)}
      +
      p_{-2}
     \mathcal{D}(-(\alpha'_{2} + \alpha'_{1}))
      e^{-i\frac{\kappa^2}{2}\sin(\Gamma_2 - \Gamma_1)} 
      \\
      &+
      p_{1}\mathcal{D}(\alpha'_{2} - \alpha'_{1})
      e^{i\frac{\kappa^2}{2}\sin(\Gamma_2 - \Gamma_1)}
      +
      p_{-1}\mathcal{D}(- (\alpha'_{2} - \alpha'_{1}))
      e^{i\frac{\kappa^2}{2}\sin(\Gamma_2 - \Gamma_1)} 
\end{align*}
and
\begin{align*}
    Q_2(\vec{\gamma}) &= \sin{\theta_{2}}\mathcal{D}(\alpha'_{2})P_1(\alpha_1) 
        -\cos{\theta_{2}}\mathcal{D}(-\alpha'_{2})Q_1(\alpha_1).
\end{align*}

It can be observed that the general form of the polynomial $P$ involve all possible summations (and subtractions) of the conditional displacement gates applied. This can be expressed as summation of sign bits multiplied with the displacement amounts resulting in several displacement gates with varying magnitude and direction. Therefore, the general form of the polynomial can be written as,

\begin{align}
    P_d(\vec{\gamma}) 
    &=
    \sum_{\textbf{s} \in \{-1, 1\}^d} p_{\textbf{s} } \mathcal{D}\left(
    \sum_{k=1}^{d}s_k\alpha'_k 
    \right)
    \exp{
    \left(
    -i\frac{\kappa^2}{2}
    \sum_{1 \leq l < k \leq d } s_k s_l
    \sin(\Gamma_k - \Gamma_l)
    \right)}, 
    \quad s_k \in \{-1, 1\}~\forall~k ~\in \{1,2, \cdots, d\}
\end{align}

\subsection{Overall Unitary for Dephasing GQSPI}
\begin{align}
    U(\beta, \kappa, \vec{\gamma}) 
     &= 
     \nonumber G^{-1}_{d,\vec{\theta}, \vec{\phi}, \lambda_0}(\vec{\gamma}) ~S_{\beta}~ G_{d,\vec{\theta}, \vec{\phi}, \lambda_0}(\vec{\gamma})
     \\
     &=  \nonumber
     \begin{bmatrix}
         P^{\dagger}_d(\vec{\gamma})& Q^{\dagger}_d(\vec{\gamma})
         \\
         * & *
     \end{bmatrix}
     R^{\dagger}_{osc}(\Gamma_d)
     ~S_{\beta}~
     R_{osc}(\Gamma_d)
     \begin{bmatrix}
         P_d(\vec{\gamma})& *
         \\
         Q_d(\vec{\gamma})& *
     \end{bmatrix}
     \\
     &= \nonumber
     \begin{bmatrix}
         P^{\dagger}_d(\vec{\gamma})& Q^{\dagger}_d(\vec{\gamma})
         \\
         * & *
     \end{bmatrix}
     e^{i\beta (\hat{p}\cos{(\Gamma_d)} - \hat{x}\sin{(\Gamma_d)})}
     \begin{bmatrix}
         P_d(\vec{\gamma})& *
         \\
         Q_d(\vec{\gamma})& *
     \end{bmatrix}
     \\
     &= \nonumber
     \begin{bmatrix}
         P^{\dagger}_d(\vec{\gamma})& Q^{\dagger}_d(\vec{\gamma})
         \\
         * & *
     \end{bmatrix}
     \mathcal{D}\left( -\frac{\beta}{\sqrt{2}}e^{i\Gamma_d} \right)
     \begin{bmatrix}
         P_d(\vec{\gamma})& *
         \\
         Q_d(\vec{\gamma})& *
     \end{bmatrix}
 \end{align}

 Since,
 \begin{align*}
     e^{i\Gamma_d \hat{n}}
     e^{i\beta \hat{p}}
     e^{-i\Gamma_d \hat{n}} &= e^{i\beta \hat{p} + \sum_{j=1}^{\infty} \frac{[i\Gamma_d \hat{n}, i\beta \hat{p}]_j}{j!}} = e^{i\beta(\hat{p}\cos{(\Gamma_d)}- \hat{x}\sin{(\Gamma_d)})}
 \end{align*}

 \begin{align*}
    i\beta (\hat{p}\cos{(\Gamma_d)} - \hat{x}\sin{(\Gamma_d)})
    &= 
    i\beta [-i\frac{1}{\sqrt{2}}( a - a^{\dagger})\cos{(\Gamma_d)}
    - \frac{1}{\sqrt{2}}(a + a^{\dagger})\sin{(\Gamma_d)}]
    \\
    \Rightarrow
    \beta_{disp} &= -\frac{\beta}{\sqrt{2}}[\cos{(\Gamma_d)} + i\sin{(\Gamma_d)}] = -\frac{\beta}{\sqrt{2}}e^{i\Gamma_d}
\end{align*}

In presence of dephasing, the signal $S_{\beta}$ undergoes rotation through an angle dependent on accumulated dephasing phase error.

The top-left element of the operator $U(\beta, \kappa, \vec{\gamma})_{00} $ in case of dephasing can be expressed as
\begin{align}
    U(\beta, \kappa, \vec{\gamma})_{00} 
    &=
    \nonumber
    \sum_{\textbf{s} \in \{-1, 1\}^d}  
    \sum_{\textbf{s'} \in \{-1, 1\}^d}
    (p^*_{\textbf{s}}p_{\textbf{s'} } + q^*_{\textbf{s} }q_{\textbf{s'} })
    \\
    & \nonumber
    \exp{
    \left(
    i\frac{\kappa^2}{2}
    \sum_{1 \leq l < k \leq d } s_k s_l
    \sin(\Gamma_k - \Gamma_l)
    \right)}
    \exp{
    \left(
    -i\frac{\kappa^2}{2}
    \sum_{1 \leq l' < k' \leq d } s'_{k'} s'_{l'}
    \sin(\Gamma_{k'} - \Gamma_{l'})
    \right)}
    \\
    & \nonumber
    \mathcal{D}^{\dagger}\left(
    \sum_{k=1}^{d}s_k\alpha'_k 
    \right)
    \mathcal{D}\left( -\frac{\beta}{\sqrt{2}}e^{i\Gamma_d} \right)
    \mathcal{D}\left(
    \sum_{k'=1}^{d}s'_{k'}\alpha'_{k'} 
    \right)
    \\
    &=
    \nonumber
    \sum_{\textbf{s}, \textbf{s'}\in \{-1, 1\}^d}  
    (p^*_{\textbf{s}}p_{\textbf{s'} } + q^*_{\textbf{s} }q_{\textbf{s'} })
    \exp{
    \left(
    i\frac{\kappa^2}{2}
    \sum_{1 \leq l < k \leq d } s_k s_l
    \sin(\Gamma_k - \Gamma_l)
    -
    i\frac{\kappa^2}{2}
    \sum_{1 \leq l' < k' \leq d } s'_{k'} s'_{l'}
    \sin(\Gamma_{k'} - \Gamma_{l'})
    \right)}
    \\
    & \label{eq:U-deph-00}
    e^{i\kappa\beta \sum_{k=1}^{d} s_k \cos{(\Gamma_d + \Gamma_k)}}
    \mathcal{D}\left( -\frac{\beta}{\sqrt{2}}e^{i\Gamma_d} \right)
    \mathcal{D}\left(
    -\sum_{k=1}^{d}s_k\alpha'_k 
    \right)
    \mathcal{D}\left(
    \sum_{k'=1}^{d}s'_{k'}\alpha'_{k'} 
    \right)
\end{align}

Here, the displacements are swapped by the relationship,
\begin{align*}
    \mathcal{D}\left(
    -\sum_{k=1}^{d}s_k\alpha'_k 
    \right)
    \mathcal{D}\left( -\frac{\beta}{\sqrt{2}}e^{i\Gamma_d} \right)
    &=
    \mathcal{D}\left( -\frac{\beta}{\sqrt{2}}e^{i\Gamma_d} \right)
    \mathcal{D}\left(
    -\sum_{k=1}^{d}s_k\alpha'_k 
    \right)
    e^{[(\frac{\beta}{\sqrt{2}}e^{i\Gamma_d})(\sum_{k=1}^{d}s_k\alpha'_k )^* - (\frac{\beta}{\sqrt{2}}e^{-i\Gamma_d})(\sum_{k=1}^{d}s_k\alpha'_k )]}
    \\
    &=
    \mathcal{D}\left( -\frac{\beta}{\sqrt{2}}e^{i\Gamma_d} \right)
    \mathcal{D}\left(
    -\sum_{k=1}^{d}s_k\alpha'_k 
    \right)
    e^{i\kappa\beta \sum_{k=1}^{d} s_k \cos{(\Gamma_d + \Gamma_k)}}
\end{align*}

with phase calculation,
\begin{align*}
    &(\frac{\beta}{\sqrt{2}}e^{i\Gamma_d})(\sum_{k=1}^{d}s_k\alpha'_k )^* - (\frac{\beta}{\sqrt{2}}e^{-i\Gamma_d})(\sum_{k=1}^{d}s_k\alpha'_k )
    \\
    &=
    (\frac{\beta}{\sqrt{2}}e^{i\Gamma_d})(\sum_{k=1}^{d}s_k \frac{\kappa}{\sqrt{2}}e^{-i(\frac{\pi}{2} - \Gamma_k)})
    - 
    (\frac{\beta}{\sqrt{2}}e^{-i\Gamma_d})
    (\sum_{k=1}^{d}s_k \frac{\kappa}{\sqrt{2}}e^{i(\frac{\pi}{2} - \Gamma_k)})
    \\
    &=
    -i\kappa\beta \sum_{k=1}^{d} s_k \sin{(\frac{\pi}{2} - \Gamma_d - \Gamma_k)}
    =
    -i\kappa\beta \sum_{k=1}^{d} s_k \cos{( - \Gamma_d - \Gamma_k)} 
    = 
    i\kappa\beta \sum_{k=1}^{d} s_k \cos{(\Gamma_d + \Gamma_k)}
\end{align*}

\subsection{\texorpdfstring{$P(M = \downarrow| \beta)$}{}}

Following a similar calculation as shown previously in Appendix~\ref{sec:prob-beta},
\begin{align*}
    P(M = \downarrow| \beta) 
    &=
    \langle 0|_{osc} U^{\dagger}(\beta, \kappa, \vec{\gamma})_{00}U(\beta, \kappa, \vec{\gamma})_{00}|0 \rangle _{osc}
    \\
    &=
    \int_{-\infty}^{\infty} 
    U^{\dagger}(\beta, \kappa, \vec{\gamma})_{00}U(\beta, \kappa, \vec{\gamma})_{00}~ \psi^{2}_0(x) dx
    \\
    &=
    \sum_{\textbf{s}, \textbf{s'}, \textbf{r}, \textbf{r'}\in \{-1, 1\}^d}  
    (p^*_{\textbf{s}}p_{\textbf{s'} } + q^*_{\textbf{s} }q_{\textbf{s'} })
    (p_{\textbf{r}}p^*_{\textbf{r'} } + q_{\textbf{r} }q^*_{\textbf{r'} })
    e^{i\Theta_{\textbf{s}, \textbf{s'}, \textbf{r}, \textbf{r'}}}
    e^{i\kappa\beta \Lambda_{\textbf{s}, \textbf{r}}}
   \left[
    \int_{-\infty}^{\infty} 
    \mathcal{D}(\alpha_{\textbf{s}, \textbf{s'}, \textbf{r}, \textbf{r'}})
    ~ \psi^{2}_0(x) dx
   \right]
   \\
   &=
   \sum_{\textbf{s}, \textbf{s'}, \textbf{r}, \textbf{r'}\in \{-1, 1\}^d}  
    (p^*_{\textbf{s}}p_{\textbf{s'} } + q^*_{\textbf{s} }q_{\textbf{s'} })
    (p_{\textbf{r}}p^*_{\textbf{r'} } + q_{\textbf{r} }q^*_{\textbf{r'} })
    e^{i\Theta_{\textbf{s}, \textbf{s'}, \textbf{r}, \textbf{r'}}}
    e^{i\kappa\beta \Lambda_{\textbf{s}, \textbf{r}}} 
    e^{-|\alpha_{\textbf{s}, \textbf{s'}, \textbf{r}, \textbf{r'}}|^2/2}
\end{align*}

where,
\begin{align*}
    &U^{\dagger}(\beta, \kappa, \vec{\gamma})_{00} U(\beta, \kappa, \vec{\gamma})_{00}
    \\
    &=
    \sum_{\textbf{s}, \textbf{s'}, \textbf{r}, \textbf{r'}\in \{-1, 1\}^d}  
    (p^*_{\textbf{s}}p_{\textbf{s'} } + q^*_{\textbf{s} }q_{\textbf{s'} })
    (p_{\textbf{r}}p^*_{\textbf{r'} } + q_{\textbf{r} }q^*_{\textbf{r'} })
    \exp{
    \left(
    i\frac{\kappa^2}{2}
    \sum_{1 \leq l < k \leq d } s_k s_l
    \sin(\Gamma_k - \Gamma_l)
    -
    i\frac{\kappa^2}{2}
    \sum_{1 \leq l' < k' \leq d } s'_{k'} s'_{l'}
    \sin(\Gamma_{k'} - \Gamma_{l'})
    \right)
    }
    \\
    &
    \exp{
    \left(
    -i\frac{\kappa^2}{2}
    \sum_{1 \leq n < m \leq d } s_m s_n
    \sin(\Gamma_m - \Gamma_n)
    +
    i\frac{\kappa^2}{2}
    \sum_{1 \leq n' < m' \leq d } s'_{m'} s'_{n'}
    \sin(\Gamma_{m'} - \Gamma_{n'})
    \right)
    }
    \\
    &
    e^{-i\kappa\beta \sum_{m=1}^{d} s_m \cos{(\Gamma_d + \Gamma_m)}}
    e^{i\kappa\beta \sum_{k=1}^{d} s_k \cos{(\Gamma_d + \Gamma_k)}}
    \\
    &
    \mathcal{D}{\dagger}\left(
    \sum_{m'=1}^{d}s'_{m'}\alpha'_{m'} 
    \right)
    \mathcal{D}{\dagger}\left(
    -\sum_{m=1}^{d}s_m\alpha'_m 
    \right)
    \mathcal{D}^{\dagger}\left( -\frac{\beta}{\sqrt{2}}e^{i\Gamma_d} \right)
    \mathcal{D}\left( -\frac{\beta}{\sqrt{2}}e^{i\Gamma_d} \right)
    \mathcal{D}\left(
    -\sum_{k=1}^{d}s_k\alpha'_k 
    \right)
    \mathcal{D}\left(
    \sum_{k'=1}^{d}s'_{k'}\alpha'_{k'} 
    \right)
    \\
    &=
    \sum_{\textbf{s}, \textbf{s'}, \textbf{r}, \textbf{r'}\in \{-1, 1\}^d}  
    (p^*_{\textbf{s}}p_{\textbf{s'} } + q^*_{\textbf{s} }q_{\textbf{s'} })
    (p_{\textbf{r}}p^*_{\textbf{r'} } + q_{\textbf{r} }q^*_{\textbf{r'} })
    e^{i\Phi_{\textbf{s}, \textbf{s'}, \textbf{r}, \textbf{r'}}}e^{i\Phi'_{\textbf{s}, \textbf{s'}, \textbf{r}, \textbf{r'}}}
    e^{i\kappa\beta \Lambda_{\textbf{s}, \textbf{r}}}
    \mathcal{D}(\alpha_{\textbf{s}, \textbf{s'}, \textbf{r}, \textbf{r'}})
    \\
    &=
    \sum_{\textbf{s}, \textbf{s'}, \textbf{r}, \textbf{r'}\in \{-1, 1\}^d}  
    (p^*_{\textbf{s}}p_{\textbf{s'} } + q^*_{\textbf{s} }q_{\textbf{s'} })
    (p_{\textbf{r}}p^*_{\textbf{r'} } + q_{\textbf{r} }q^*_{\textbf{r'} })
    e^{i\Theta_{\textbf{s}, \textbf{s'}, \textbf{r}, \textbf{r'}}}
    e^{i\kappa\beta \Lambda_{\textbf{s}, \textbf{r}}} 
    \mathcal{D}(\alpha_{\textbf{s}, \textbf{s'}, \textbf{r}, \textbf{r'}}) 
    \quad\quad 
    \text{where, }
    (\Theta_{\textbf{s}, \textbf{s'}, \textbf{r}, \textbf{r'}} = \Phi_{\textbf{s}, \textbf{s'}, \textbf{r}, \textbf{r'}}+\Phi'_{\textbf{s}, \textbf{s'}, \textbf{r}, \textbf{r'}})
\end{align*}

The expressions for the factors are:
\begin{align}
    e^{i\Phi_{\textbf{s}, \textbf{s'}, \textbf{r}, \textbf{r'}}}
    &= \nonumber
    \exp{
    \left(
    i\frac{\kappa^2}{2} \left[
    \sum_{1 \leq l < k \leq d } s_k s_l
    \sin(\Gamma_k - \Gamma_l)
    -
    \sum_{1 \leq l' < k' \leq d } s'_{k'} s'_{l'}
    \sin(\Gamma_{k'} - \Gamma_{l'})
    -
    \sum_{1 \leq n < m \leq d } r_m r_n
    \sin(\Gamma_m - \Gamma_n)
     \right. \right. }
    \\
    &{
    \left. \left.
    +
    \sum_{1 \leq n' < m' \leq d } r'_{m'} r'_{n'}
    \sin(\Gamma_{m'} - \Gamma_{n'})
    \right]
    \right)
    }
\end{align}
\begin{align}
    e^{i\Phi'_{\textbf{s}, \textbf{s'}, \textbf{r}, \textbf{r'}}}
    &=  \nonumber
    \exp{ \Biggl[i\frac{\kappa^2}{2}\left( 
    \sum_{m',m = 1}^{d} r'_{m'}r_m\sin{(\Gamma_{m'} - \Gamma_{m})}
    -
    \sum_{m', k = 1}^{d} r'_{m'}s_k\sin{(\Gamma_{m'} - \Gamma_{k})}
    +
    \sum_{m', k' = 1}^{d} r'_{m'}s_{k'}\sin{(\Gamma_{m'} - \Gamma_{k'})}
    \right. 
    }
    \\
    &{
    \left.
    + \sum_{m, k = 1}^{d} r_{m}s_k\sin{(\Gamma_{m} - \Gamma_{k})}
    -
    \sum_{m, k' = 1}^{d} r_{m}s'_{k'}\sin{(\Gamma_{m} - \Gamma_{k'})}
    +
    \sum_{k,k' = 1}^{d} s_{k}s'_{k'}\sin{(\Gamma_{k} - \Gamma_{k'})}
    \right) \Biggr]} 
\end{align}

\begin{align}
    e^{i\kappa\beta \Lambda_{\textbf{s}, \textbf{r}}} 
    =
    e^{i\kappa\beta [\sum_{k=1}^{d} s_k\cos{(\Gamma_d + \Gamma_k)} - \sum_{m=1}^{d}r_m\cos{(\Gamma_d + \Gamma_m)}]}
\end{align}

\begin{align}
    \mathcal{D}(\alpha_{\textbf{s}, \textbf{s'}, \textbf{r}, \textbf{r'}})
    &=
    \mathcal{D}\left(
    -\sum_{m'=1}^{d}r'_{m'}\alpha'_{m'} 
    +
    \sum_{m=1}^{d}r_m\alpha'_m 
    -
    \sum_{k=1}^{d}s_k\alpha'_k 
    +
    \sum_{k'=1}^{d}s'_{k'}\alpha'_{k'} 
    \right)
\end{align}

These factors would simplify further as many of the arguments either add up or cancel out based on the composition of the displacement gates.
\subsection{Effect of small \texorpdfstring{$\gamma$}{} on \texorpdfstring{$P(M = \downarrow| \beta)$}{}}

For small $\gamma$, we can approximate 
\begin{align*}
    sin{(a\gamma)} \approx a\gamma - \frac{a^3\gamma^3}{6} + O(\gamma^5), \quad \cos{(a\gamma)} \approx  1 - \frac{a^2\gamma^2}{2} + O(\gamma^4)  
\end{align*}

We first approximate each factor in $P(M = \downarrow| \beta)$. 

On computing the argument of $e^{i\Theta_{\textbf{s}, \textbf{s'}, \textbf{r}, \textbf{r'}}}$, it can be observed that it would consist of a summation of sinusoids with some coefficients. Thus,
\begin{align}
    e^{i\Theta_{\textbf{s}, \textbf{s'}, \textbf{r}, \textbf{r'}}
     }
     & \nonumber
     \equiv e^{i\frac{\kappa^2}{2}\sum_a A_a \sin{a\gamma}}, \quad a \in \mathbb{Z}, A_a \in \mathbb{R}
     \\ 
     &\approx e^{i\frac{\kappa^2}{2} (A \gamma+ A' \gamma^3)}, \quad A, A' \in \mathbb{R}
\end{align}

Similarly, since $s_k$ and $r_m$ are both sign values $\pm 1$, the cosines with same argument either add up or cancel out.
\begin{align}
    e^{i\kappa\beta \Lambda_{\textbf{s}, \textbf{r}}} 
    &= \nonumber
    e^{i\kappa\beta [\sum_{k=1}^{d} s_k\cos{(\Gamma_d + \Gamma_k)} - \sum_{m=1}^{d}r_m\cos{(\Gamma_d + \Gamma_m)}]}
    \equiv e^{i2\kappa\beta \sum_b \cos{(b\gamma})}, \quad b \in \mathbb{Z}
    \\
    \nonumber
    &
    \approx e^{i2\kappa\beta B}e^{-i2\kappa\beta B'\gamma^2}
    \quad B, B' \in \mathbb{R}
\end{align}

\begin{align*}
    \mathcal{D}(\alpha_{\textbf{s}, \textbf{s'}, \textbf{r}, \textbf{r'}})
    &\approx 
    \mathcal{D}\left(\sum_c C_c \frac{\kappa}{\sqrt{2}} e^{i(\frac{\pi}{2} - c\gamma)}\right)
    \approx \mathcal{D}
    \left(
    \sum_c C_c \frac{\kappa}{\sqrt{2}} 
    (i\cos{c\gamma} + \sin{c\gamma})
    \right),\quad  C_c \in \mathbb{R}
    \\
    &\approx
    \mathcal{D}
    \left(
     \frac{\kappa}{\sqrt{2}} 
    [i(C - C'\gamma^2) + C''\gamma]
    \right), \quad\quad C,C',C'' \in \mathbb{R}
    \\
  e^{-|\alpha_{\textbf{s}, \textbf{s'}, \textbf{r}, \textbf{r'}}|^2/2}  
  &\equiv e^{-(D + D'\gamma^4 + D''\gamma^2)/2}\quad\quad D, D', D'' \in \mathbb{R}
\end{align*}

\begin{align*}
    &P(M = \downarrow| \beta) 
   =
   \sum_{\textbf{s}, \textbf{s'}, \textbf{r}, \textbf{r'}\in \{-1, 1\}^d}  
    (p^*_{\textbf{s}}p_{\textbf{s'} } + q^*_{\textbf{s} }q_{\textbf{s'} })
    (p_{\textbf{r}}p^*_{\textbf{r'} } + q_{\textbf{r} }q^*_{\textbf{r'} })
    e^{i(\Theta_{\textbf{s}, \textbf{s'}, \textbf{r}, \textbf{r'}}
     + \Phi_{\textbf{s}, \textbf{s'}, \textbf{r}, \textbf{r'}})}
    e^{i\kappa\beta \Lambda_{\textbf{s}, \textbf{r}}} 
    e^{-|\alpha_{\textbf{s}, \textbf{s'}, \textbf{r}, \textbf{r'}}|^2/2}
    \\
    &=
    \sum_{\textbf{s}, \textbf{s'}, \textbf{r}, \textbf{r'}\in \{-1, 1\}^d}  
    (p^*_{\textbf{s}}p_{\textbf{s'} } + q^*_{\textbf{s} }q_{\textbf{s'} })
    (p_{\textbf{r}}p^*_{\textbf{r'} } + q_{\textbf{r} }q^*_{\textbf{r'} })
    e^{i\frac{\kappa^2}{2} (A \gamma+ A' \gamma^3)}
    e^{i2\kappa\beta B}e^{-i2\kappa\beta B'\gamma^2}
    e^{-\kappa^2(D + D'\gamma^4 + D''\gamma^2)/2}
    \\
    &\approx
    \sum_{\textbf{s}, \textbf{s'}, \textbf{r}, \textbf{r'}\in \{-1, 1\}^d}  
    (p^*_{\textbf{s}}p_{\textbf{s'} } + q^*_{\textbf{s} }q_{\textbf{s'} })
    (p_{\textbf{r}}p^*_{\textbf{r'} } + q_{\textbf{r} }q^*_{\textbf{r'} })
    [1+i\frac{\kappa^2}{2} (A \gamma+ A' \gamma^3) - \frac{\kappa^4}{4} (A \gamma+ A' \gamma^3)^2]
    \\
    &
    (1-i2\kappa\beta B'\gamma^2 - 2\kappa^2\beta^2 B'^2\gamma^4)
    [1-\frac{\kappa^2}{2}(D'\gamma^4 + D''\gamma^2)]e^{i2\kappa\beta B}e^{-D\frac{\kappa^2}{2}}
    \\
    &\approx
    \sum_{\textbf{s}, \textbf{s'}, \textbf{r}, \textbf{r'}\in \{-1, 1\}^d}  
    (p^*_{\textbf{s}}p_{\textbf{s'} } + q^*_{\textbf{s} }q_{\textbf{s'} })
    (p_{\textbf{r}}p^*_{\textbf{r'} } + q_{\textbf{r} }q^*_{\textbf{r'} })
    [1+iA_1 \gamma + A_2\gamma^2 +iA_3 \gamma^3 + O(\gamma^4)](1-iB_1\beta\gamma^2 - O(\gamma^4)), 
    \quad A_1, A_2, A_3, B_1, C_1 \in \mathbb{R}
    \\
    &
    (1- C_1\gamma^2 - O(\gamma^4))e^{i2\kappa\beta B}e^{-D\frac{\kappa^2}{2}}
    \\
    &\approx
     \sum_{\textbf{s}, \textbf{s'}, \textbf{r}, \textbf{r'}\in \{-1, 1\}^d}  
    (p^*_{\textbf{s}}p_{\textbf{s'} } + q^*_{\textbf{s} }q_{\textbf{s'} })
    (p_{\textbf{r}}p^*_{\textbf{r'} } + q_{\textbf{r} }q^*_{\textbf{r'} }) 
    e^{i2\kappa\beta B}
    e^{-D\frac{\kappa^2}{2}}
    [1 - \Omega_1 \gamma^2 - \Omega_2\beta \gamma^3 + i(\Omega_3 \gamma + \Omega_4 \gamma^3)+\mathcal{O}(\gamma^4)],
    \quad \Omega_1, \Omega_2\in\mathbb{R}
\end{align*}

Since the probability is real-valued, the linear imaginary term must cancel. Consequently, the probability response function exhibits damping due to the second- and third-order terms in $\gamma$, along with slight deviations arising from the $\beta$-dependent third-order term as $\beta$ moves further away from $\beta = 0$ and the threshold values.

\section{Frame Analysis for Signal Detection of Squeezed Signals 
}
\label{appndx:squeezed-signals-frame}

\subsection{Family of Frame Functions}
Consider the family of functions defined in Eq.~\eqref{eq:frame-unit},
which arise in the representation of $P(M=\downarrow \mid \beta)$. We assume that the parameter $\beta$ lies within a finite interval $[\mathcal{B}_-, \mathcal{B}_+]$, rather than the entire real line, and the functions $g_{s,l}$ are square-integrable in this interval, i.e., $g_{s,l} \in L^2([\mathcal{B}_-, \mathcal{B}_+])$.

For our case, $\mathcal{B}_- = -\frac{\pi}{2\kappa}$ and  $\mathcal{B}_+ = \frac{\pi}{2\kappa}$, due to the periodicity of the protocol as discussed in Sec.~\ref{sec:mp_ta-gqspi-thm}. 
Therefore we only consider the interval of interest to be $L^2([-\frac{\pi}{2\kappa}, \frac{\pi}{2\kappa}])$.

For $s=0$, the functions reduce to
\begin{align}
    g_{0,l}(\beta) = 1, \quad \forall l,
\end{align}
which is a point, and implies that the family contains repeated elements. Since $g_{0,l}(\beta)$ does not add to the span of the family of functions, we remove it to eliminate redundancy. Further, it can be observed that,
\begin{align}
    g_{s,l}(\beta) =  g_{-s,-l}(\beta)
\end{align}
since the ranges of $l$ and $s$ are symmetric with respect to positive and negative indices. Thus, we define a reduced index set
\begin{align}
    \mathcal{I}_{\mathrm{red}} = \{(0,0)\} \cup \{(s,l):  s > 0 \},
\end{align}
and consider the reduced family of function $\{g_{s,l}(\beta)\}_{(s,l)\in \mathcal{I}_{\mathrm{red}}}$. The reduced set now only contains distinctly modulated Gaussian window functions.


For a given $s$, the width of the Gaussian window in the $e^{-4s^2\kappa^2\beta^2}$ term becomes fixed for all indices of $l$. Then, the term $e^{i(4\kappa^2\beta s)l}$ is essentially a Fourier frame \cite{1d493e2c-cfa9-3224-9c3d-d3d9555183b1, 6723d674-1f20-39d1-a0fa-adb4f9f0358a} on $L^2([-\frac{\pi}{2\kappa}, \frac{\pi}{2\kappa}])$ with identically weighted windows. Assume we fix $s$ to be an integer $s_0> 0$. Then,
\begin{align}
    g_{s_0,l}(\beta) = e^{i(4\kappa^2s_0)l\beta } e^{-(4s_0^2\kappa^2)\beta^2}.
\end{align}
Here, the weighting factor $w(\beta) = e^{-(4s_0^2\kappa^2)\beta^2}$ satisfies, 
\begin{align}
   e^{-s_0^2 \pi^2} \le w(\beta) \le 1,
\end{align}
since,
\begin{align}
    |\beta| < \frac{\pi}{2\kappa}.
\end{align}
Because multiplying a known frame by a bounded, nonzero weight also produces a frame, the sub-family of functions $g_{s_0,l}(\beta)$ form a frame on $L^2([-\frac{\pi}{2\kappa}, \frac{\pi}{2\kappa}])$. Further, since this is true for all $s \ne 0$, then their superposition as shown in Eq.~\eqref{eq:sup-sub-fam-frame} also forms a frame due to linearity
\begin{align}
    \sum_{s > 0}^{d} \sum_{l = -2d}^{2d}\alpha_{s,l} e^{i(4\kappa^2s)l\beta } e^{-(4s^2\kappa^2)\beta^2} 
    \label{eq:sup-sub-fam-frame}.
\end{align}
\subsection{Conditions on $\kappa$}
Since the span of the frames is $L^2([-\frac{\pi}{2\kappa}, \frac{\pi}{2\kappa}])$, we must obtain the conditions for the bounds on $\kappa > 0$ for which the system forms a frame. Given that $\beta \in (-\frac{\pi}{2\kappa}, \frac{\pi}{2\kappa})$, therefore, the factor 
 $4\kappa^2 s_0\beta $ belongs to the interval,
\begin{align}
    4\kappa^2 s_0\beta \in (-2\pi \kappa s_0, 2\pi \kappa s_0)
\end{align}
which has length, $4\pi \kappa s_0$. However, Fourier frames have a length of $2\pi$, therefore,
\begin{align}
    &4\pi \kappa s_0 \le 2\pi \nonumber
    \\
    &\Rightarrow \kappa  \le \frac{1}{2 s_0}.
\end{align} 
The aforementioned condition must hold for all $s_0$, we set the value of $s = d$ which is the highest index of $|s|$ in the protocol. Thus, the following condition,
\begin{align}
    \kappa  \le \frac{1}{2d}
\end{align}
is required for the family of functions represented in Eq.~\eqref{eq:frame-unit} to form a Fourier frame. This also implies that, the higher the degree $d$ of the protocol, the smaller displacement steps are involved in performing such multi-resolution interrogation on the squeezed signal.


\end{document}